\documentclass[a4paper]{article}

\usepackage{amssymb}
\usepackage{amsmath}
\usepackage{graphicx}

\usepackage{array}
\newcolumntype{x}[1]{>{\centering\arraybackslash\hspace{0pt}}p{#1}}
\usepackage{amsthm}
\usepackage{epstopdf}
\usepackage{graphics}
\usepackage{graphicx}
\usepackage{latexsym}
\usepackage[margin=1in]{geometry}

\usepackage[utf8]{inputenc}
\usepackage{hyperref}
\usepackage{multirow}
\usepackage{float}
\usepackage{subfig}

\usepackage{algorithmicx}
\usepackage{algorithm}
\usepackage{lscape}
\usepackage[noend]{algpseudocode}

\usepackage[numbers]{natbib}

\usepackage{xcolor}
\hypersetup{
    colorlinks,
    linkcolor={red!50!black},
    citecolor={blue!50!black},
    urlcolor={blue!80!black}
}
\title{Performance Analysis of Metaheuristic Optimization Algorithms in Estimating the Interfacial Heat Transfer Coefficient on Directional Solidification}

\author{Gianfranco de M. Stieven$^1$\\
	Erb F. Lins$^{1,*}$ \\
	Edilma P. Oliveira$^2$ \\
	\small $^{1}$ Faculty of Mechanical Engineering, Federal University of Par\'a \\
	\small Av. Augusto Correa, Nº 1 - Bel\'em, PA, 66075-900, Brazil \\
	\small $^{2}$Faculty of Mechanical Engineering, Federal University of South and Southeast of Par\'a \\
	\small Folha 17, Quadra 04, Lote Especial, Nova Marab\'a, Marab\'a, PA, 68505-080, Brazil \\
	\small $^{*}$ Corresponding author: E. Lins (erb@ufpa.br)
}

\date{\today} 

\providecommand{\keywords}[1]
{
	\small	
	\textbf{\textit{Keywords---}} #1
}

\begin{document}

\maketitle

\begin{abstract}
In this paper is proposed an evaluation of ten metaheuristic optimization algorithms applied on the inverse optimization of the Interfacial Heat Transfer Coefficient (IHTC) coupled on the solidification phenomenon. It was considered an upward directional solidification system for Al-7wt.\% Si alloy and, for IHTC model, a exponential time function. All thermophysical properties of the alloy were considered constant. Scheil Rule was used as segregation model ahead phase-transformation interface. Optimization results from Markov Chain Monte Carlo method (MCMC) were considered as reference. Based on average, quantiles 95\% and 5\%, kurtosis, average iterations and absolute errors of the metaheuristic methods, in relation to MCMC results, the Flower Pollination Algorithm (FPA) and Moth-Flame Optimization (MFO) presented the most appropriate results, outperforming the other methods in this particular phenomenon, based on these metrics. The regions with the most probable values for parameters in IHTC time function were also determined.
\end{abstract}


\section{Introduction}
\label{intro}

Many problems in science deal with parameters uncertainty. As a result, predictions of nature's behavior may lose accuracy and scientific conclusions may be compromised. To overcome this problem several algorithms, called optimization methods, have been proposed. These computational strategies play an important role: parameter estimation using numerical routines derived from differential operators, bayesian methodologies or even nature observations techniques. In the last case, studies involving the so-called Nature-Inspired Optimization Algorithms (NIOA) and Metaheuristic Optimization Algorithms (MOA) are usual. These methods are based on a local and global random search and use weighting mechanisms adapted from the behavior of, for example, living beings, celestial bodies and natural phenomena. In this group of algorithms, it is worth pointing out widely used ones to predict parameters, such as Simulated Annealing \cite{kirkpatrick1983optimization}, Ant Colony Optimization \cite{dorigo1999ant, lakshmanaprabu2019effect}, Cuckoo Search \cite{yang2009cuckoo, chi2019hybridization}, Wind Driven Algorithm \cite{bayraktar2013wind, abdalla2019wind}, Tree-Seed Algorithm \cite{kiran2015tsa, becskirlicomparison}, Water Wave Optimization \cite{zheng2015water, shao2019novel}, Tree-Growth Algorithm \cite{cheraghalipour2018tree}, Sailfish Optimizer \cite{shadravan2019sailfish} and Manta Ray Foraging Optimization \cite{zhao2020manta}. Due to easy of implementation, exploration and exploitation search, convergence speed and operational versatility, these algorithms have become popular in parameter estimation in several applications, such as crack propagation \cite{cheng2018control}, design of heat exchangers \cite{de2019design}, photovoltaic systems \cite{tey2018implementation}, power control of wind turbine \cite{benamor2019new}, strategic aircraft deconfliction \cite{courchelle2019simulated}, optimal power flow problems \cite{abdullah2019optimal}, hybrid renewable systems \cite{samy2019flower} and others.

By No Free Lunch (NFL) theorems \cite{yang2018mathematical, joyce2018review}, originally established for search in \cite{wolpert1995no} and for optimization in \cite{wolpert1997no}, it is known that no optimization algorithm is better than others over all possible functions and real-world problems. However, ss reminded by \cite{mcdermott2019and}, despite of some incorrect ideas about NFL, the original statement does not ignore the fact that there is some algorithm $A$ that can outperform $B$, once $A$ could be specialized to the set of problems analyzed in the case. NFL encourages researchers to identify this specialization \cite{yang2012free}. One can do that by creating a tailored-hand algorithm to the problem to achieve better than random performance or identifying among all developed algorithms one that is already specialized to some or a particular problem subset.
So, it is necessary to analyze case-by-case the performance of metaheuristic methods to define, in a restrictive way to the physical model, which method outperforms others in terms of parameter estimation, processing time, operational cost and other quantitative and qualitative requirements.

In this context, there has been little information available about the effectiveness of optimization metaheuristic methods in specific applications of Mechanical Engineering and Metallurgy. Even though in many papers the optimization of test functions is approached \cite{xue2018self, ezugwu2019conceptual}, there is also great utility in verifying the numerical performance of the optimization methods in more complex, real-world problems. One of these is the transient heat and mass transfer coupled with phase transformation, that is, the phenomenon of solidification on permanent metallic mold castings. In this process, the Interfacial Heat Transfer Coefficient (IHTC) represents one of the most important thermal coefficients since it predominantly controls the heat transfer of the metallic system, then directly influencing several thermal parameters. These can be explicitly related to the physical \cite{santos2017, satbhai2017} and mechanical \cite{spin2012, brito2015} metallurgy of the ingot. Although its relevance, this coefficient is of difficult estimation since it is not easy to measure experimentally and comes from a numerical ill-posed problem. In this way, it is necessary to use an optimization method that, from the experimental data, can infer this parameter from an inverse approach.

So, in this paper, ten optimization methods will be analyzed in order to determine which of them, in a solidification problem, returns the best temperature prediction, which are: Particle Swarm Optimization \cite{kennedy1995particle}, Differential Evolution \cite{storn1997}, Bat Algorithm  \cite{yang2010}, Flower Pollination Algorithm \cite{yang2012}, Grey Wolf Optimizer \cite{mirjalili2014grey}, Moth-Flame Optimization \cite{mirjalili2015moth}, Sine Cosine Algorithm \cite{mirjalili2016sca}, Whale Optimization Algorithm \cite{mirjalili2016whale}, Dragonfly Algorithm \cite{mirjalili2016dragonfly} and Harris Hawks Optimization \cite{heidari2019harris}. These methods were chosen based on their scientific relevance and applicability, once these methods have already been applied by several researchers in the most varied areas of science and their performances are notoriously recognized.

Markov Chain Monte Carlo (MCMC) method will be used in order to determine the statistical consistency of the results provided by the optimization algorithms under study.  This method is widely used for estimations in materials and thermal science, such as thermal diffusivity of metals \cite{gnanasekaran2013markov}, heat flux \cite{kumar2018bayesian}, heat transfer coefficient \cite{szenasi2018using} and metallic fatigue \cite{babuvska2016bayesian}.

Therefore, the main contribution of this paper is a performance analysis of various metaheuristic algorithms on the inverse optimization of the IHTC in directional solidification. Results obtained via Markov Chain Monte Carlo will be considered, in order to determine, given the physical phenomenon, which method stands for the most appropriated for inference.

An aluminum-based alloy was the material chosen to be analyzed. This type of alloy has a notorious importance, since, in an effort to improve vehicle fuel efficiency by  lower consumption, lightweight aluminum alloys have been replacing other materials for use in automotive integrated systems and components \cite{johnson2020development} such as engine blocks and cylinder heads \cite{stroh2019effects} and also on aerospace and marine applications \cite{kordijazi2020statistical}. Among aluminum-based alloys, the aluminum-silicon (Al-Si) system has considerable prestige. Al-Si alloys are widely used in the automotive industry, especially on engines \cite{chen2020effect}, due to their attractive strength to weight ratio as well as superior casting characteristics \cite{rakhmonov2017influence}. Regarding the production volume,
Al-Si alloys account for
80 to 90\% of the aluminum castings produced
commercially \cite{ASM2018_1}. Based on this author, there is an optimum range of silicon content, depending on the casting process:
5 to 7wt.\% Si for slow cooling-rate processes,
such as casting in plaster, investment,
and sand molds; 7 to 9wt. \% Si for permanent molds;
and 8 to 12wt. \% Si for die casting.

Therefore, given the industrial and academic relevance of this non-ferrous system and its usage ranges, this paper chose Al-7wt.\% Si (also registered by the Aluminum Association (AA) under the number A444.0) as the study alloy. This alloy, especially on permanent mold casting, exhibit excellent proprieties such as resistance to hot cracks, corrosion resistance and good fluidity, shrinkage tendency, castability and weldability \cite{ASM2018_2}. 

The remainder of this paper is organized as follows. Section \ref{model} describes the numerical methodology, in which the Finite Volume Method and boundary conditions are presented. In Section \ref{metaheuristics} is expressed the optimization strategies contained on the metaheuristic optimizers applied in this paper. After, Section \ref{mcmc} presents a brief statistical review of MCMC, stochastic method that will be used as reference for this performance analysis. In Section \ref{methodology} is explained the heat transfer inverse methodology applied on parameter estimation via MOA and MCMC. Section \ref{results_discussion} holds the contributions of this paper, including MCMC analysis, evaluation of MOA, comparison of experimental and simulated thermal profiles, convergence, error analysis and posterior probability distribution for the IHTC parameters. Finally, Section \ref{conclusion} concludes the paper by presenting the overall results obtained in this contribution.

\section{Mathematical and Numerical Descriptions for Direct Problem}
\label{model}

In this Section an introduction is done in order to present the heat problem in analysis. We consider a numerical model based on Finite Volume Method. In this technique, the domain is divided into a set of interconnected volumes and, afterwards, the conservation law is applied to each volume. More details of the mathematical approach can be seen in \cite{edilma2019}.  

To model the phenomenon of solidification, the transient heat transfer equation for an arbitrary volume $\Omega $, bounded by a surface $S $, is used,

\begin{equation}
\label{eq1}
\int_{\Omega}{\rho c\frac{\partial T}{\partial t}d\Omega} + \int_{S}{\mathbf{F}\cdot d\mathbf{A}} = \int_\Omega {\dot{q} d\Omega} 
\end{equation}
where $T\equiv T(\mathbf{z},t)$ represents temperature, \(\mathbf{z}\) is a spatial vector, $t$ is time, $\rho~\equiv~\rho(\mathbf{z},t)$ represents the material density and $c\equiv c(\mathbf{z},t)$ the specific heat. $\mathbf{F}$ is the heat flux through the boundary surface $S $, $d\mathbf{A}$ the elemental area projected on the surface $S$ and $\dot {q}$ the internal energy generated or consumed within the control volume due to the energy sources. Here, the contribution of the internal energy $\dot {q}$ is only by latent heat, which is defined as,
\begin{equation}
\label{eq2}
\dot {q}=\rho L\frac{\partial f_s }{\partial t}
\end{equation}
where $L$ is the latent heat coefficient and $f_s$ the local solid fraction. To estimate $f_s$ in the mushy zone, Scheil's rule was used (which is well used as segregation model ahead phase-transformation interface), given by Eq. (\ref{scheil}),
\begin{equation}
\label{scheil}
f_s = 1 - \left(\frac{T_f - T}{T_f - T_{liq}} \right)^{\frac{1}{k_0-1}}
\end{equation} where \(T_f\) represents the melting temperature of the pure solvent, \(T_{liq}\) the \textit{liquidus} temperature of the metal alloy and \(k_0\) the partition coefficient. The derivative $\frac{\partial f_s }{\partial t}$ in Eq. \ref{eq2} may be computed as a function of $\frac{\partial T}{\partial t}$ using a pseudo specific heat,
\begin{equation}
\label{eq3}
c_p =c_m -L\frac{\partial f_s }{\partial T}
\end{equation}
where $c_m $ is the specific heat at mushy zone, which is given by a simple mixing law as,
\begin{equation}
c_m =f_s c_s +(1-f_s )c_l
\end{equation}
where $c_s $ and $c_l $ are the specific heats of solid and liquid zones, respectively. Thermal conductivity and density are given by,
\begin{eqnarray}
k &=& f_s k_s +(1-f_s )k_l \\
\rho &=&  f_s \rho_s +(1-f_s ) \rho_l
\end{eqnarray}
where $k_s $ and $k_l $ are thermal conductivity of solid and liquid zones, as well as $\rho_s $ and $ \rho_l $ are density of solid and liquid zones, respectively. All thermophysical properties from the alloy are considered constant.  Supposing an upward solidification device, the boundary conditions for the mathematical approach are: the lateral walls of the mold are thermal insulated, and the heat is extracted only from the lower boundary, so that solidification occurs only on vertical upward direction. The convection effect is not considered, once its effect is very low in upward solidification \cite{wang2018}.

To represent the IHTC, the model expressed in Eq. \ref{atb} \cite{el1991, zeng2015, edilma2019, vishweshwara2019} was chosen, once is one of the most applied non-constant model of IHTC. In this equation, $t$ represents time and \(t_0\) a referential time (\(t_0\) = 1 s) and $A$ \(\left[\frac{\text{W}}{\text{m}^2 \text{K}}\right]\) and $B$ \(\left[-\right]\) constants.

\begin{equation}
h_i = A \left( \dfrac{t}{t_0} \right)^{B}
\label{atb}
\end{equation}

\section{Metaheuristic Optimization Algorithms}
\label{metaheuristics}

This Section presents a description of the ten metaheuristic methods applied in this paper. For primary source information and details of each algorithm, see \cite{kennedy1995particle, storn1996usage, storn1997, yang2010, yang2012, mirjalili2014grey, yang2014nature, mirjalili2015moth, mirjalili2016sca, mirjalili2016whale, mirjalili2016dragonfly, heidari2019harris}.

\subsection{Particle Swarm Optimization}
\label{sub:PSO}

The Particle Swarm Optimization (PSO) method was developed by \cite{kennedy1995particle} in 1995, which was inspired by the observation of natural clusters, such as migratory behavior of fish and birds \cite{yang2014nature}. It was created as an alternative to the genetic algorithm that, until that time, was restricted to modeling with binaries. This method is based on the social behavior of various species, trying to balance the individuality and sociability of individuals in order to find an optimal interest \cite{orlande2011thermal}.

The PSO method is convenient because it does not use the differential concept, performing heuristic optimization only by consulting the objective function and weighting particles. Thus, the method is in many cases superior in processing and simulation time than other heuristic methods such as genetic algorithm and simulated annealing \cite{hanrahan2011swarm}.

In this algorithm, consider \(\mathbf{X} = x^t_i  \) and \(\mathbf{V} = v^t_i\), respectively, the position and velocity for particle $i$ at generation or iteration $t$. The particles \(x^{t}_{*}  \) and \(g_*\) represent the best position on the iteration $t$ and the best ones obtained so far on the simulation, respectively. The velocity upgrade is determined by Eq. \ref{PSO_v},where \(\theta_{\text{PSO}}\) represents the inertia constant, \(\alpha_{\text{PSO}}\) and \(\beta_{\text{PSO}}\) the acceleration constants and \(u_1\) and \(u_2\) random numbers uniformly distributed, such that \(u_{\Box} = \text{U}\left[ 0,1 \right] \). The initial velocity of the particle can be zero, that is, \(v_{i}^{0} = 0 \). The new particle position is updated as Eq. \ref{PSO_x}.
\begin{equation}
v^{t+1}_{i} = \theta_{\text{PSO}} \cdotp v^{t}_{i} + \alpha_{\text{PSO}} \cdot u_1 \cdotp \left[ g_{*} - x^{t}_{i} \right] + \beta_{\text{PSO}} \cdotp u_2 \cdotp \left[x^{t}_{*} - x^{t}_{i} \right] \label{PSO_v}
\end{equation}
\begin{equation}
x^{t+1}_{i} = x^{t}_{i} + v^{t+1}_{i} \label{PSO_x}
\end{equation}

In Algorithm \ref{alg_PSO} a pseudo-code of this method is presented. The mathematical explanation of the equations below was taken from \cite{yang2014nature}.

\begin{algorithm} [H]
	\caption{Pseudo-code of PSO algorithm \cite{yang2014nature}}
	\begin{algorithmic}
		\State Objective function \( \mathcal{F}\left(\mathbf{X}\right), \mathbf{X} = x^t_i\)
		\State Initialize locations \(\mathbf{X}\) and velocity \(\mathbf{V}\) of \(n_{\text{part}} \), where \(n_{\text{part}} \) is the number of particles 
		\State Find \( g_* \) for min\( \left[\mathcal{F}(\mathbf{X}) \right] (\text{at $t$ = 0}) \)
		\While{(criterion)}
		\For{each search agent}
		\State Generate new velocity \(v^{t+1}_{i}\) using Eq. \ref{PSO_v} 
		\State Calculate new locations using Eq. \ref{PSO_x}
		\State Evaluate objective functions at new locations \(x^{t+1}_{i}\)
		\State Find the current best for each particle \(x^{t}_{*}\)
		\EndFor
		\State \textbf{end for}
		\State Find the current global best \(g_{*}\)
		\EndWhile
		\State \textbf{end while}
		\State Output the final results \(x^{t}_*\) and \(g_{*}\)
	\end{algorithmic}
	\label{alg_PSO}
\end{algorithm}

\subsection{Differential Evolution}
\label{sub:DE}

Differential Evolution (DE) is a vector-based algorithm created by \cite{storn1996usage, storn1997} to minimize nonlinear and non-differentiable continuous space functions. This method consists in mutation, crossover and selection. 
Consider \(\mathbf{X} = x^t_i  \) and \(\mathbf{V} = v^t_i\), respectively, the position and velocity for particle $i$ at generation or iteration $t$. A velocity \(v^t_i\), generated by randomly selected \(x^t_i\) positions, here established as \(x^{t}_{k_{i}}\), is used for mutation. In this paper, \(\mathbf{V}\) is created by a DE/Rand/2/Bin scheme \cite{yang2014nature}, which includes five \(x^{t}_{k_{i}}\) positions, upgrading mutation, as can be seen in Eq. \ref{DE_v}. In this equation, \(F^1_{\text{DE}} \) and \(F^2_{\text{DE}}\) are differential weights, such that  \(F^{\Box}_{\text{DE}} \in \left[0,2\right]\).

\begin{equation}
v_{i}^{t+1} = x^{t}_{k_{1}} + F^1_{\text{DE}} \cdotp  \left(x^{t}_{k_{2}} - x^{t}_{k_{3}}\right) + F^2_{\text{DE}} \cdotp \left(x^{t}_{k_{4}} - x^{t}_{k_{5}}\right) \label{DE_v}
\end{equation}

After mutation, crossover is used to select randomly if \(v_{i}^t\) or \(x^{t}_{i}\) will be used in the next step of optimization. That decision is based on a crossover parameter \(C_{r_{\text{DE}}} \in \left[0,1\right]\), as expressed in Eq. \ref{DE_u}. In this equation, $u_1$ represents a uniformly distributed random number, such that \(u_{\Box} = \text{U}\left[ 0,1 \right]\).

\begin{equation}
u^{t+1}_{i} = \left\{\begin{matrix}
v^{t}_{i}, \ \ \ \text{if}  \ u_1 \leq C_{r_{\text{DE}}} \\ \\
x_{i}^{t}, \ \ \ \ \ \ \ \text{otherwise}
\end{matrix} \right.
\label{DE_u}
\end{equation}

At last, in selection, the position \( u_{i}^{t+1} \) and \( x^{t}_{i} \) will be compared in terms of the direct problem output for each vector \((\mathcal{F}(\Box))\), as in Eq. \ref{DE_x}.

\begin{equation}
x^{t+1}_{i} = \left\{\begin{matrix}
u_{i}^{t+1}, & \text{if} \ \ \ \mathcal{F}\left(u_{i}^{t+1}\right) \leq \mathcal{F}\left(x_{i}^{t}\right), \\ \\
x_{i}^{t}, & \text{otherwise}.
\end{matrix} \right.
\label{DE_x}
\end{equation}

This process continues until convergence. The principal steps to implement DE can be seen on Algorithm \ref{alg_DE}.

\begin{algorithm} [H]
	\caption{Pseudo-code of DE algorithm \cite{yang2014nature}}  
	\begin{algorithmic}
		\State Initialize the population \(\mathbf{X}\) with randomly generated solutions
		\State Set the weights \(F^1_{\text{DE}} \) and \(F^2_{\text{DE}} \in \left[0,2 \right] \) and crossover probability \(C_{r_{\text{DE}}} \in \left[0,1 \right] \)
		\While{(criterion)}
		\For{each search agent}
		\State Select randomly \(x^t_{k_1}\), \(x^t_{k_2}\), \(x^t_{k_3}\), \(x^t_{k_4}\) and \(x^t_{k_5}\)
		\State Generate velocity \(\mathbf{V}\) by Eq. \ref{DE_v} 
		\State Generate an uniformly distributed random number \(u_1 = \text{U}\left[0,1\right] \)
		\State Select and update the solution by Eq. \ref{DE_u} and \ref{DE_x} 
		\EndFor		
		\State \textbf{end for}
		\EndWhile
		\State \textbf{end while}
		\State Post-process and output the best solution found
	\end{algorithmic}
	\label{alg_DE}
\end{algorithm}

\subsection{Bat Algorithm}
\label{sub:BA}

Bat Algorithm (BA) was created by \cite{yang2010} in 2010. This method is inspired on the behavior of bats, especially by their capacity of echolocation. Based on \cite{yang2014nature}, the algorithm follows the approximate rules:

\begin{enumerate}
	\item All bats use echolocation to sense distance, and they also ``know" the difference between food/prey and background barriers;
	\item Bats fly randomly with velocity \(v_{i}^t\) at position \(x_{i}^t\). They can automatically adjust the frequency of their emitted pulses and adjust the rate of pulse emission \(r_{\text{BA}} \in \left[0,1\right]\), depending on the proximity of their target;
	\item Although the loudness can vary in many ways, we assume that the loudness \(\varphi_{\text{BA}_{i}}\) varies from a large (positive) \(\varphi_{\text{BA}_{0}}\) to a minimum value \(\varphi_{\text{BA}_{min}}\).
\end{enumerate}

Defining the rules of how position \(\mathbf{X}\) and velocities \(\mathbf{V}\) are updated, Eq. \ref{BA_f}, \ref{BA_v} and \ref{BA_x} express the numerical update procedure. In these expressions, \(f_{\text{BA}} \in \left[f_{\text{BA}_{min}}, f_{\text{BA}_{max}}\right]\) is the frequency value, \(u_{\Box} = \text{U}\left[0,1\right] \) is an uniformly distribution random number and \(x^{t}_{*}\) is the current best global location or solution.

\begin{equation}
f_{\text{BA}_{i}} = f_{\text{BA}_{min}} + u_1\cdotp\left( f_{\text{BA}_{max}} - f_{\text{BA}_{min}} \right)
\label{BA_f}
\end{equation}

\begin{equation}
v^{t+1}_{i} = v^{t}_{i} + f_{\text{BA}_{i}}  \cdotp  \left(x^{t}_{i} - x^{t}_{*} \right) 
\label{BA_v}
\end{equation}

\begin{equation}
x^{t+1}_{i} = x^{t}_{i} + v^{t+1}_{i} \label{BA_x}
\end{equation}

After this computation, an uniformly distributed random number \(u_{\Box}\) is generated. If it is greater than the pulse rate \(r_{\text{BA}}\), \(x_{i}^{t+1}\) suffers a perturbation as in Eq. \ref{BA_xnew}, where \(\varphi_{\text{BA}_{mean}}\) is the average loudness of all the bats at this time step, \(n_{\Box} = \text{N}\left(0,1\right)\) is a normally distributed random number and \(\sigma_{\text{BA}}\) is a scaling factor.

\begin{equation}
x_{i}^{t+1} = x_{i}^{t}+ \sigma_{\text{BA}} \cdotp n_1 \cdotp \varphi_{\text{BA}_{mean}}
\label{BA_xnew}
\end{equation}

After this procedure, an \(u_{\Box}\) is selected again. Two conditions need to be confirmed: if \(u_{\Box}\) is lower than the loudness \(\varphi_{\text{BA}_{i}}\) and the fitness \(\mathcal{F}\left( x^t_{i} \right) \) is lower than the fitness \(\mathcal{F}\left(x^{t}_{*}\right)\), the new point is accept as the new solution and the pulse rate and loudness suffer alterations, as expressed in Eq. \ref{BA_at+1}. Is assumed that \(\varphi_{\text{BA}_{0}} = 1\), \(\varphi_{\text{BA}_{min}} = 0\), \(\gamma_{\text{BA}}\) and \(\alpha_{\text{BA}}\) are constants and \(k_{int}\) represents iteration. For better understanding on Algorithm \ref{alg_BA} is summarized all the steps to code this method.

\begin{equation}
\varphi_{\text{BA}_{i}}^{t+1} = \alpha_{\text{BA}} \varphi_{\text{BA}_{i}}^{t}, \ \ \ r^{t+1}_{\text{BA}_{i}} = \varphi_{\text{BA}_{i}}^{0} \left[1 - \text{exp}\left(- \gamma_{\text{BA}} k_{int}\right)\right]
\label{BA_at+1}
\end{equation}

\begin{algorithm} [H]
	\caption{Pseudo-code of BA \cite{yang2014nature}}
	\begin{algorithmic}
		\State Initialize \(\mathbf{X}\) and \(\mathbf{V}\) 
		\State Initialize \(f_{\text{BA}_{i}}\), pulse rate \(r_{\text{BA}_{i}}\) and loudness \(\varphi_{\text{BA}_{i}}\)
		\While{(criterion)}
		\State Generate new solutions by adjusting frequency
		\State Update velocities and locations/solutions by Eq. \ref{BA_f}, \ref{BA_v} and \ref{BA_x}
		\If{\(u_1 > r_{\text{BA}_{i}}\)} 
		\State Select a solution among the best solutions 
		\State Generate a local solution around the selected best solution by Eq. \ref{BA_xnew}
		\EndIf
		\State \textbf{end if}
		\State Generate a new solution by flying randomly
		\If {\(u_2 < \varphi_{\text{BA}_{i}}\) and \(\mathcal{F}(x^t_{i}) < \mathcal{F}(x^{t}_{*} )\)}
		\State Accept the new solution
		\State Increase \(r_{\text{BA}_{i}}\) and reduce \(\varphi_{\text{BA}_{i}}\) by Eq. \ref{BA_at+1}
		\EndIf
		\State \textbf{end if}		
		\State Rank the bats and find the current best \(x^{t}_*\)
		\EndWhile
		\State \textbf{end while}
	\end{algorithmic}
	\label{alg_BA}
\end{algorithm}

\subsection{Flower Pollination Algorithm}
\label{sub:FPA}

Flower Pollination Algorithm (FPA) was created by \cite{yang2012} in 2012 based on the behavior of the flow pollination process of flowering plants. As written in \cite{yang2014nature}, this method can be explained by four rules:

\begin{enumerate}
	\item Biotic and cross-pollination can be considered processes of global pollination, and pollen-carrying pollinators move in a way that obeys L\'evy flights;
	\item For local  pollination, abiotic pollination and self-pollination are used;
	\item Pollinators such as insects can develop flower constancy, which is equivalent to a reproduction probability that is proportional to the similarity of two flowers involved;
	\item The interaction or switching of local and global pollination can be controlled by a switch probability \(p_{\text{FPA}} \in \left[0,1\right]\), slightly biased toward local pollination.
\end{enumerate}

Mathematically, rule 1 and 3 can be represented as in Eq. \ref{FPA_x_1}, where \(\mathbf{X} = x^{t}_{i}\) is the pollen (particle) $i$ at iteration $t$, \(x^{t}_*\) is the current best solution found among all solutions, \(\gamma_{\text{FPA}}\) is a scaling factor, \(L_{\text{FPA}}\) is a step-size parameter, which can be observed in Eq. \ref{FPA_L}, and \(\lambda_{\text{FPA}}\) is a step-size constant. In this equation, \(\Gamma\) represents the Gamma Function.

\begin{equation}
x^{t+1}_{i} = x^{t}_{i} + \gamma_{\text{FPA}} \cdotp L_{\text{FPA}}\left(\lambda_{\text{FPA}}\right) \cdotp \left(x^{t}_* - x^{t}_{i}\right)
\label{FPA_x_1}
\end{equation}

\begin{equation}
L_{\text{FPA}}\left(\lambda_{\text{FPA}}\right) = \dfrac{\lambda_{\text{FPA}} \cdotp \Gamma\left(\lambda_{\text{FPA}}\right) \cdotp \text{sin}\left(\frac{\pi \cdotp \lambda_{\text{FPA}}}{2}\right)}{\pi} \cdotp \dfrac{1}{\left(s_{\text{FPA}}\right)^{1+\lambda_{\text{FPA}}}}
\label{FPA_L}
\end{equation}

The step-size $s_{\text{FPA}}$ can be expressed by Eq. \ref{FPA_s}, which relates \(\lambda_{\text{FPA}}\) and two Gaussian distributions: \(n_g = \text{N}\left(0, \sigma^2_{\text{FPA}}\right)\) and \(n_1 = \text{N}\left(0, 1\right)\). The variance \(\sigma^2_{\text{FPA}}\) can be calculated by Eq. \ref{FPA_sigma2}.

\begin{equation}
s_{\text{FPA}} = \dfrac{n_g}{|n_1|^{\frac{1}{\lambda_{\text{FPA}}}}}
\label{FPA_s}
\end{equation}

\begin{equation}
\sigma^2_{\text{FPA}} = \left[\dfrac{\Gamma\left(1+\lambda_{\text{FPA}}\right)}{\lambda_{\text{FPA}} \cdotp \Gamma \left(\frac{1 + \lambda_{\text{FPA}}}{2}\right)} \cdotp \dfrac{\text{sin}\left(\frac{\pi \cdotp \lambda_{\text{FPA}}}{2}\right)}{2^{\frac{\lambda_{\text{FPA}} - 1}{2}}}     \right]^{\frac{1}{\lambda_{\text{FPA}}}}
\label{FPA_sigma2}
\end{equation}

For local pollination, rules 2 and 3 can be mathematically expressed as in Eq. \ref{FPA_x_2}, where \(x^{t}_{j}\) and \(x^{t}_{k}\) are pollen from different flower of the same plant species. In this equation, \(u_1 = \text{U}\left[0,1\right] \).

\begin{equation}
x^{t+1}_{i} = x^{t}_{i} + u_1 \cdotp \left(x^{t}_{j} - x^{t}_{k}\right)
\label{FPA_x_2}
\end{equation}

For better understanding of this method, Algorithm \ref{alg_FPA} explains by a pseudo-code its numerical implementation.

\begin{algorithm} [H]
	\caption{Pseudo-code of FPA \cite{yang2014nature}}
	\begin{algorithmic}
		\State Objective min or max \( \mathcal{F}\left(\mathbf{X}\right)\), \(\mathbf{X} = x^t_i \) 
		\State Initialize a population of $n_{\text{part}}$ flower/pollen gametes with random solutions
		\State Find the best solution \(x^{*}_i\) in the initial population
		\State Define a switch probability \(p_{\text{FPA}} \in \left [ 0,1 \right ] \)
		\While{(criterion)}
		\For{each search agent}
		\If{$u_{\Box} = \text{U}\left[0,1\right] < p_{\text{FPA}}$}
		\State Global pollination via  Eq. \ref{FPA_x_1} 
		\Else
		\State Local pollination via Eq. \ref{FPA_x_2} 
		\EndIf
		\State \textbf{end if}
		\State Evaluate new solutions
		\State If new solutions are better, update them in the population
		\EndFor
		\State \textbf{end for}
		\State Find the current best solution \(x^{t}_*\)
		\EndWhile
		\State \textbf{end while}
		\State Output the best solution found
	\end{algorithmic}
	\label{alg_FPA}
\end{algorithm}

\subsection{Grey Wolf Optimizer}
\label{sub:GWO}

The Grey Wolf Optimizer (GWO) method was conceived by \cite{mirjalili2014grey} in 2014, which were inspired on hunting technique and the social hierarchy of grey wolves. The understanding of this method requires a piece of information of how to manipulate mathematically the behavior of these animals. Comparing the social hierarchy of wolves and optimization language, the alpha wolf \((\alpha_w)\) is considered the fittest solution. The second and third best solutions are named beta \(\left(\beta_w\right)\) and delta \(\left(\delta_w\right)\). The rest of the solutions is named omega \(\left(\omega_w\right)\). In this method, \(\alpha_w\), \(\beta_w\) and \(\delta_w\) wolves are the hunters and \(\omega_w\) wolves follow them.

The mathematical approach for encircling prey are expressed by Eq. \ref{GWO_xt+1}, where \(A_{\text{GWO}}\) (in Eq. \ref{GWO_A}) and \(C_{\text{GWO}}\) (in Eq. \ref{GWO_C}) are coefficients, \(x^t_{p}\) is the position of the prey, \(x^t_i\) is the position of a grey wolf, \(a_{int}\) is a linearly decreasing parameter that goes from 2 to 0 based on the number of iterations and \(u_1\) and \(u_2\) are uniformly distributed random numbers.

\begin{equation}
x^t_i = x_p^t - A_{\text{GWO}} \cdotp |C_{\text{GWO}} \cdotp x _p^t- x^t_i|
\label{GWO_xt+1}
\end{equation}

\begin{equation}
A_{\text{GWO}} = 2 \cdotp a_{int} \cdotp u_1 - a_{int}
\label{GWO_A}
\end{equation}

\begin{equation}
C_{\text{GWO}} = 2 \cdotp u_2
\label{GWO_C}
\end{equation}

As the analogy suggests, because we do not know the solution to the problem, that is, the position of the prey, the new position suggested by the formulation of this method is obtained from the position of the $\alpha_w$, $\beta_w$ and $\delta_w$ wolves. Replacing $x^t_p$ by $x^t_i$, it is suggested that the location of the prey (optimal point) is close to the location of the wolves (points with the best position until the current iteration). The Equations \ref{GWO_XI}, \ref{GWO_XII} and \ref{GWO_XIII} expose this behavior, where \(x^t_{\text{I}}\), \(x^t_{\text{II}}\) and \(x^t_{\text{III}}\) represent, respectively, the best, second and third best positions, guided by the wolves position, that is, 
\(x^t_{\alpha_w}\), \(x^t_{\beta_w}\) and \(x^t_{\delta_w}\). The new position is determined by the arithmetic mean of the best positions constructed by the location of the dominant lobes (best positions previously obtained), as expressed in Eq. \ref{GWO_Xs/3}. 


\begin{equation}
x^t_{\text{I}} = x^t_{\alpha_w} -  A_{\text{GWO}} \cdotp
|C_{\text{GWO}} \cdotp x^t_{\alpha_w} - x^t_i|
\label{GWO_XI}
\end{equation}

\begin{equation}
x^t_{\text{II}} = x^t_{\beta_w} -  A_{\text{GWO}} \cdotp
|C_{\text{GWO}} \cdotp x^t_{\beta_w} - x^t_i|
\label{GWO_XII}
\end{equation}

\begin{equation}
x^t_{\text{III}} = x^t_{\delta_w} -  A_{\text{GWO}} \cdotp
|C_{\text{GWO}} \cdotp x^t_{\delta_w} - x^t_i|
\label{GWO_XIII}
\end{equation}

\begin{equation}
x^{t+1}_i = \dfrac{x^t_{\text{I}} + x^t_{\text{II}} + x^t_{\text{III}}}{3}
\label{GWO_Xs/3}
\end{equation}

On Algorithm \ref{alg_GWO} it is possible to observe a pseudo-code containing the programming procedure of this method based on the formulations presented above.

\begin{algorithm} [H]
	\caption{Pseudo-code of GWO algorithm \cite{mirjalili2014grey}}
	\begin{algorithmic}
		\State Initialize the grey wolf population \(\mathbf{X} \)
		\State Initialize $a_{\text{GWO}}$, $A_{\text{GWO}}$ and $A_{\text{GWO}}$
		\State Calculate the fitness of each search agent
		\State Identify \(x^t_{\alpha_w}, x^t_{\beta_w}\) and \(x^t_{\delta_w}\) (first, second and third best solutions)
		\While{(criterion)}
		\For{each search agent}
		\State Update the position of the current search agent by Eq. \ref{GWO_Xs/3}
		\EndFor
		\State \textbf{end for}
		\State Update $a_{int}$, $A_{\text{GWO}}$ and $C_{\text{GWO}}$
		\State Calculate the fitness of all search agent
		\State Update \(x^t_{\alpha_w}, x^t_{\beta_w}\) and \(x^t_{\delta_w}\)
		\State \(t = t + 1\)
		\EndWhile
		\State \textbf{end while}
	\end{algorithmic}
	\label{alg_GWO}
\end{algorithm}

\subsection{Moth-Flame Optimization}
\label{sub:MFO}

The Moth-Flame Optimization (MFO) algorithm was conceived by \cite{mirjalili2015moth} in 2015, which was inspired on the navigation method of moths in nature called transverse orientation.

Consider \(\mathbf{M} = m_i^t\) and \(\mathbf{F} = f_i^t\) the moth and flame positions, respectively, in which $i$ and $t$ are, respectively, the particle and iteration index. \(\mathbf{OM} = om_i^t\) and \(\mathbf{OF} = of_{i}^t\) represents the fitness values of the moth and flame particles, such that \(\mathbf{OM} = \mathcal{F}\left(\mathbf{M}\right) \) and \(\mathbf{OF} = \mathcal{F}\left(\mathbf{F}\right) \). As exposed in \cite{mirjalili2015moth}, it is worth pointing out that moths and flames are both solutions. The difference between them is the way we treat and update them in each iteration. The moths are actual search agents that move around the search space, whereas flames are the best position of moths that obtains so far. In other words, $\mathbf{F}$ is a sorted vector obtained by the best particles in $\mathbf{M}$ during the entire simulation. Flames, in this method, are considered as flags that are dropped by moths when searching the search space. Equation \ref{MFO_3.11} represents the modification of the moth position with respect to the geometry \( \left( \mathcal{S\left(\Box\right)}  \right) \) used to search the best position.

\begin{equation}
m_{i}^{t+1} = \mathcal{S} \left(m_i^t, f_i^t\right)
\label{MFO_3.11}
\end{equation} 

In the original paper a logarithmic spiral was chosen, which formulation can be seen in Eq. \ref{MFO_3.12}, where \(d_i^t\) indicates the distance between moth and flame (Eq. \ref{MFO_3.13}), $b_{\text{MFO}}$ is a constant for defining the shape of the logarithmic spiral, and $t_{int}$ is a random number in $[-1, 1]$.

\begin{equation}
\mathcal{S} \left(m_i^t, f^t_i\right) = d_i^t \cdotp \text{e}^{\left(b_{\text{MFO}} \cdotp t_{int}\right)} \cdotp \text{cos}\left(2\cdotp \pi \cdotp t_{int}\right) + f_i^t
\label{MFO_3.12}
\end{equation} 

\begin{equation}
d^t_i = |f_i^t - m_i^t|
\label{MFO_3.13}
\end{equation}

In order to balance exploration and exploitation during the iterative process, it was observed that the number of flames should be reduced in order to favor exploitation in the search for more promising solutions. Thus, a formulation was proposed with the objective of reducing the number of search positions in a given iteration number, which can be seen in Eq. \ref{MFO_3.14},

\begin{equation}
f_n = \left \lfloor N_{max} - t \cdotp \dfrac{N_{max} - 1}{T_{int}} \right \rceil
\label{MFO_3.14}
\end{equation} where $t$ is the current number of iteration and $N_{max}$ and $T_{int}$ represent the maximum number of flames and iterations, respectively. In Equation \ref{MFO_3.14}, \(\left \lfloor  \Box \right \rceil\) represents rounding to the nearest integer, such that \(  \left \lfloor  \Box \right \rceil  = \left \lfloor \Box + 0.5  \right \rfloor \).

A pseudo-code containing the programming procedure of this method based can be seen on Algorithm \ref{alg_MFO}.

\begin{algorithm} [H]
	\caption{Pseudo-code of MFO algorithm \cite{mirjalili2015moth}}
	\begin{algorithmic}
		\State Update $f_n$ by Eq. \ref{MFO_3.14}
		\State Create \(\mathbf{M}\) and \(\mathbf{OM}\)
		\If{iteration == 1}
		\State \(\mathbf{F}\) = \text{sort}(\(\mathbf{M}\))
		\State \(\mathbf{OF}\) = \text{sort}(\(\mathbf{OM}\))
		\Else
		\State \(\mathbf{F}\) = \text{sort}(\(\mathbf{M}_{t-1}, \mathbf{M}_{t}\))
		\State \(\mathbf{OF}\) = \text{sort}(\(\mathbf{M}_{t-1}, \mathbf{M}_{t}\))
		\EndIf
		\State \textbf{end if}
		\While{(criterion)}
		\For{each search agent}
		\State Calculate $d_i^t$ using Eq. \ref{MFO_3.13} with respect to the corresponding moth
		\State Update $m^t_i$ using Eqs. \ref{MFO_3.11} and \ref{MFO_3.12} with respect to the corresponding moth 
		\EndFor
		\State \textbf{end for}
		\EndWhile
		\State \textbf{end while}
	\end{algorithmic}
	\label{alg_MFO}
\end{algorithm}

\subsection{Sine Cosine Algorithm}
\label{sub:SCA}

Sine Cosine Algorithm (SCA) is population-based optimization algorithm published by \cite{mirjalili2016sca} in 2016 that uses a mathematical model based on sine and cosine functions for solving optimization problems. This method is easy to implement because it uses only two search functions, which are changeable according to a random number. The construction of this algorithm, based on the following equations, can be observed in Algorithm \ref{alg_SCA}. 

Consider \(\mathbf{X} = x^{t}_{i}\) the $i$-th position of the current solution at $t$-th iteration. In Equation \ref{SCA_3.3} the random exchange mechanism between the two search equations is shown. In this equation, \(r_{iter}\) is a parameter given by Eq. \ref{SCA_3.4}, \(r^1_{\text{U}}\) and \(r^2_{\text{U}}\) are random numbers such that \(r^1_{\text{U}} = 2 \pi  u_1\) and \(r^2_{\text{U}} = 2 u_2\), in which \(u_1\) and \(u_2\) are uniformly distributed random numbers \(\left(u_{\Box} = \text{U} \left[0,1\right]   \right)  \) and \(g_*\) is the best position obtained so far.

\begin{equation}
x^{t+1}_{i} =  \left\{\begin{matrix}
x^{t+1}_{i} = x^{t}_{i} + r_{int} \cdotp  \text{sin}\left(r^1_{\text{U}}\right) \cdot |r^2_{\text{U}} \cdotp g_{*} - x^{t}_{i}|, \ \ \ \text{if} \ \ u_3 < 0.5
\\ 
x^{t+1}_{i} = x^{t}_{i} + r_{int} \cdotp  \text{cos}\left(r^1_{\text{U}}\right) \cdot |r^2_{\text{U}} \cdotp g_{*} - x^{t}_{i}|, \ \ \ \text{if} \ \ u_3 \ge 0.5
\end{matrix} \right.
\label{SCA_3.3}
\end{equation}

 In Equation \ref{SCA_3.4}, $t$ is the current iteration, $k_{max}$ is the maximum number of iterations and $a_{\text{SCA}}$ is a constant.

%

\begin{equation}
r_{int} = a_{\text{SCA}} - t \cdotp \dfrac{a_{\text{SCA}}}{k_{max}}
\label{SCA_3.4}
\end{equation}

\begin{algorithm} [H]
	\caption{Pseudo-code of SCA \cite{mirjalili2016sca}}
	\begin{algorithmic}
		\State Initialize a set of search agents \(\mathbf{X}\)
		\While{(criterion)}
		\State Evaluate each of the search agents by the objective function 
		\State Update the position of search agents using Eq. \ref{SCA_3.3} and \ref{SCA_3.4}
		\EndWhile
		\State \textbf{end while}
	\end{algorithmic}
	\label{alg_SCA}
\end{algorithm}

\subsection{Whale Optimization Algorithm}
\label{sub:WOA}

Whale Optimization Algorithm (WOA) was developed by \cite{mirjalili2016whale} in 2016. This optimizer mimics the social behavior of humpback whales. As expressed in the original paper, whales can recognize the location of prey and encircle them. Since the position of the optimal design in the search space is not known a priori, the WOA algorithm assumes that the current best candidate solution is the target prey or is close to the optimum, which is a similar strategy applied in Section \ref{sub:GWO} and is represented by Eq. \ref{WOA_2.2}. In this equation, \(A_{\text{WOA}}\) (Eq. \ref{WOA_2.3}) and \(C_{\text{WOA}}\) (Eq. \ref{WOA_2.4}) are coefficients, \(a_{int}\) is linearly decreased parameter from 2 to 0 over the course of iterations, $u_1$ and $u_2$ is an uniformly distributed random number, such that \(u_{\Box} = \text{U}\left[0,1\right] \), \(g_{*}\) is the best solution obtained so far and \(\mathbf{X} = x^t_i\) is the particle position.

\begin{equation}
x^{t+1}_i = g_* - A_{\text{WOA}} \cdotp |C_{\text{WOA}} \cdotp g_* - x^t_i|
\label{WOA_2.2}
\end{equation} 
\begin{equation}
A_{\text{WOA}} = 2 \cdotp a_{int} \cdotp u_1 - a_{int}
\label{WOA_2.3}
\end{equation}
\begin{equation}
C_{\text{WOA}} = 2 \cdotp u_2
\label{WOA_2.4}
\end{equation}

For exploitation, Eq. \ref{WOA_2.5} is used to perform a spiral updating position, which mimics the helix-shaped movement of humpback whales, very similar to the one presented in Section \ref{sub:MFO},
\begin{equation}
x^{t+1}_i = d_i^t \cdotp \text{e}^{\left(b_{\text{WOA}} \cdotp t_{int}\right)} \cdotp \text{cos}\left(2 \cdotp \pi \cdotp t_{int}\right) + g_*
\label{WOA_2.5}
\end{equation} where \(d_i^t = |g_* - x^t_i|\) indicates the distance of the $i$-th position to the best solution obtained so far, $b_{\text{WOA}}$ is a constant related to the shape of the spiral and $t_{int}$ is a random number in $[-1, 1]$.

Two mechanisms are presented simultaneously by humpback whales: swim around the prey within a shrinking circle and along a spiral-shaped path. In the original paper is assumed probability of 50\% for occurrence of both models, as expressed in Eq. \ref{WOA_2.6}.

\begin{equation}
x^{t+1}_i =  \left\{\begin{matrix}
g_* - A_{\text{WOA}} \cdotp |C_{\text{WOA}} \cdotp g_* - x_i^t| , \ \ \ \text{if} \ \ u_3 < 0.5
\\ 
d_i^t \cdotp \text{e}^{\left(b_{\text{WOA}} \cdotp t_{int}\right)} \cdotp \text{cos}\left(2 \cdotp \pi \cdotp t_{int}\right) + g_*, \ \ \ \text{if} \ \ u_3 \ge 0.5
\end{matrix} \right.
\label{WOA_2.6}
\end{equation}

According to the behavior of \(A_{\text{WOA}}\), the prey search equations are altered in order to favor exploitation, as seen in Eq. \ref{WOA_2.8}. This search modification can be observed on Algorithm \ref{alg_WOA}.

\begin{equation}
x_i^{t+1} = x_{k_1}^t - A_{\text{WOA}} \cdotp |C_{\text{WOA}} \cdotp  x_{k_1}^t - x^t_i|
\label{WOA_2.8}
\end{equation} where \(x_{k_i}^t\) is a random position chosen from the current population.

\begin{algorithm} [H]
	\caption{Pseudo-code of WOA \cite{mirjalili2016whale}}
	\begin{algorithmic}
		\State Initialize the whales population \(\mathbf{X}\)
		\State Calculate the fitness of each search agent
		\State \(g_{*}\) = the best search agent
		\While{(criterion)}
		\For{earch search agent}
		\State Update $a_{int}$, $A_{\text{WOA}}$, $C_{\text{WOA}}$, $t_{int}$ and $u_3$
		\If{\(u_3 < 0.5\)}
		\If{\(|A_{\text{WOA}}| < 1\)}
		\State Update the position of the current search agent by the Eq. \ref{WOA_2.2}
		\ElsIf{\(|A_{\text{WOA}}| \geq 1\)}
		\State Update the position of the current search agent by the Eq. \ref{WOA_2.8}
		\EndIf
		\State \textbf{end if}
		\ElsIf{\(u_3 \geq 0.5\)}
		\State Update the position of the current search by the Eq. \ref{WOA_2.5}
		\EndIf
		\State \textbf{end if}	
		\EndFor
		\State \textbf{end for}
		\State Check if any search agent goes beyond the search space and amend it
		\State Calculate the fitness of each search agent
		\State Update \(g_{*} \) if there is a better solution
		\State \(t = t + 1\) 
		\EndWhile
		\State \textbf{end while}
		\State Return \(g_{*}\)
	\end{algorithmic}
	\label{alg_WOA}
\end{algorithm}

\subsection{Dragonfly Algorithm}
\label{sub:DA}

Dragonfly Algorithm (DA) was created by \cite{mirjalili2016dragonfly} in 2016. This method is inspired on the static and dynamic swarming behaviors of dragonflies in nature. As expressed on the original paper, based on \cite{reynolds1987flocks}, the behavior of swarms follows three primitive principles:

\begin{enumerate}
	\item Separation: Static collision avoidance
	of the individuals from other individuals in the
	neighbourhood;
	\item Alignment: Velocity matching of individuals
	to that of other individuals in neighbourhood;
	\item Cohesion: Tendency of individuals
	towards the center of the mass of the neighbourhood.
\end{enumerate}

So, the principal goal for any swarm is survival, so all of the individuals should be attracted towards food sources and distracted outward enemies. Mathematically, separation is calculated by Eq. \ref{DA_3.1}.

\begin{equation}
S_i^{\text{DA}} = - \sum^{N_{\text{nb}}}_{j=1} \left(x_i^t - x_j^t\right)
\label{DA_3.1}
\end{equation} where $x_i^t$ is the position of the current individual, \(x_j^t\) is the position $j$-th neighboring individual and $N_{\text{nb}}$ is the number of neighboring individuals. Alignment is calculated as in Eq. \ref{DA_3.2}, where \(v_j^t\) represents the velocity of $j$-th neighboring individual.

\begin{equation}
A_i^{\text{DA}} = \dfrac{\sum_{j=1}^{N_{nb}}v_j^t} {N_{nb}}
\label{DA_3.2}
\end{equation}

Cohesion is calculated by Eq. \ref{DA_3.3}, where \(x^t_i\) is the position of the current individual and \(x_j^t\) shows the position $j$-th neighboring individual.

\begin{equation}
C_i^{\text{DA}} = \dfrac{\sum_{j=1}^{N_{nb}}x_j^t} {N_{nb}} - x^t_i
\label{DA_3.3}
\end{equation}

Attraction towards a food source \(\left(F_i^{\text{DA}} \right)  \) is calculated as
Eq. \ref{DA_3.4}, where \(x^{+}_i\) shows the position of the food source. Distraction outwards an enemy \(\left(E_i^{\text{DA}} \right)  \) is calculated as Eq. \ref{DA_3.5}, where \(x^{-}_i\) shows the position of the enemy.

\begin{equation}
F_i^{\text{DA}}  = x^{+}_i - x^t_i
\label{DA_3.4}
\end{equation} 

\begin{equation}
E_i^{\text{DA}} = x^{-}_i - x_i^t
\label{DA_3.5}
\end{equation} 

The step $\Delta x^{t+1}_i$ shows the movement direction of the dragonflies, which is defined in Eq. \ref{DA_3.6}, where $s_{\text{DA}}$ shows the separation weight, \(S_i^{\text{DA}}\) indicates the separation of the $i$-th individual, $a_{\text{DA}}$ is the alignment weight, \(A_i^{\text{DA}}\) is the alignment of $i$-th individual, $c_{\text{DA}}$ indicates the cohesion weight, \(C_i^{\text{DA}}\) is the cohesion of the $i$-th individual, $f_{\text{DA}}$ is the food factor, \(F_i^{\text{DA}}\) is the food source of the $i$-th individual,
$e_{\text{DA}}$ is the enemy factor, \(E_i^{\text{DA}}\) is the position of enemy of the $i$-th individual and $w_{\text{DA}}$ is the inertia weight. After that, the position vector are updated as Eq. \ref{DA_3.7}.

\begin{equation}
\Delta x^{t+1}_i = \left(s_{\text{DA}} \cdotp S_i^{\text{DA}} + a_{\text{DA}} \cdotp A_i^{\text{DA}} + c_{\text{DA}} \cdotp C_i^{\text{DA}} + f_{\text{DA}} \cdotp F_i^{\text{DA}} + e_{\text{DA}} \cdotp E_i^{\text{DA}}\right) + w_{\text{DA}} \cdotp \Delta x^t_i
\label{DA_3.6}
\end{equation}

\begin{equation}
x^{t+1}_i = x^{t}_i + \Delta x^{t+1}_i
\label{DA_3.7}
\end{equation}

One mechanism to enhance the stochastic behavior and exploration of the particles is by adding a random walk by L\'evy flight where there is no neighboring solutions. So, the position of dragonflies can be updated by Eq. \ref{DA_3.8}. The L\'evy flight can be calculated by Eq. \ref{DA_3.9}, where \(u_1\) and \(u_2\) are uniformly distributed random numbers in $[0, 1]$, \(\lambda_{\text{DA}}\) is a constant and \(\sigma_{\text{DA}}\) is calculated by Eq. \ref{DA_3.10}, where \(\Gamma\) represents the Gamma Function.

\begin{equation}
x^{t+1}_i = x^{t}_i + L_{\text{DA}}\left(\lambda_{\text{DA}} \right) \cdotp x^{t}_i
\label{DA_3.8}
\end{equation}

\begin{equation}
L_{\text{DA}}\left(\lambda_{\text{DA}} \right) = 0.01 \cdotp \dfrac{u_1 \cdotp \sigma_{\text{DA}}}{|u_2|^{\frac{1}{\lambda_{\text{DA}}}}}
\label{DA_3.9}
\end{equation}

\begin{equation}
\sigma_{\text{DA}} = \left[\dfrac{\Gamma\left(1+\lambda_{\text{DA}}\right)}{\lambda_{\text{DA}} \cdotp \Gamma \left(\frac{1 + \lambda_{\text{DA}}}{2}\right)} \cdotp \dfrac{\text{sin}\left(\frac{\pi\cdotp\lambda_{\text{DA}}}{2}\right)}{2^{\frac{\beta - 1}{2}}}     \right]^{\frac{1}{\lambda_{\text{DA}}}}
\label{DA_3.10}
\end{equation}

A better understanding of the code building process can be obtained from the pseudo-code expressed on Algorithm \ref{alg_DA}.

\begin{algorithm} [H]
	\caption{Pseudo-code of DA \cite{mirjalili2016dragonfly}}
	\begin{algorithmic}
		\State Initialize the dragonflies population \(x_i^t \)
		\State Initialize step \(\Delta x^t_i\)
		\While{(criterion)}
		\State Calculate the objective values of all dragonflies
		\State Update the food source and enemy
		\State Update $w_{\text{DA}}$, $s_{\text{DA}}$, $a_{\text{DA}}$, $c_{\text{DA}}$, $f_{\text{DA}}$ and $e_{\text{DA}}$
		\State Calculate $S_i^{\text{DA}}$, $A_i^{\text{DA}}$, $C_i^{\text{DA}}$, $F_i^{\text{DA}}$ and $E_i^{\text{DA}}$ using Eqs. \ref{DA_3.1} to \ref{DA_3.5}
		\State Update neighbouring radius
		\If{a dragonfly has at least one neightbouring dragonfly}
		\State Update velocity  using Eq. \ref{DA_3.6}  
		\State Update position using Eq. \ref{DA_3.7}
		\Else
		\State Update position using Eq. \ref{DA_3.8}
		\EndIf
		\State \textbf{end if}
		\State Check and correct the new positions based on the boundaries of variables
		\EndWhile
		\State \textbf{end while}
	\end{algorithmic}
	\label{alg_DA}
\end{algorithm}

\subsection{Harris Hawks Optimization}
\label{sub:HHO}

Harris Hawks Optimization (HHO) was developed by \cite{heidari2019harris} in 2019, which were inspired on the behavior and chasing style of Harris hawks in nature called ``surprise pounce".

In this method, as expressed in the original paper, the Harris hawks are the candidate solutions and the best candidate solution in each step is considered as the intended prey or nearly the optimum. The Harris' hawks perch randomly on some locations and wait to detect a prey based on two strategies, resulting on an equal chance for each perching strategy. They perch based on the positions of other family members and the rabbit, which is modeled in Eq. \ref{HHO_1}, where \(x^{t+1}_i\) is the position of hawks in the next iteration, \(x^t_*\) is the position of rabbit, that is, the best solution in the iteration $t$, \(x^t_i\) is the current position vector of hawks, \(u_1\), \(u_2\), \(u_3\), \(u_4\) and \(u_5\) are uniformly distributed random numbers \(\left(u_{\Box} = \text{U}\left(0,1\right)   \right) \), $LB$ and $UB$ show the upper and lower bounds of the variable, \(x_{k_i}^t\) is a randomly selected hawk from the current population and \(x^t_{mean}\) is the average position of the current population of hawks. In Equation \ref{HHO_2}, $n_{\text{part}}$ denotes the total number of hawks.

\begin{equation}
x^{t+1}_i=  \left\{\begin{matrix}
x_{k_1}^{t} - u_1 \cdotp |x_{k_1}^{t} - 2 \cdotp u_2 \cdotp  x^t_i| , \ \ \ \ \ \ \ \ \ \ \ \ \ \ \ \ \ \ \ \ \ \ \ \ \ \ \ \text{if} \ \ u_5 \ge 0.5
\\ 
\left(x^t_{k_i} - x^t_{mean}\right) - u_3 \cdotp \left[LB + u_4 \cdotp\left(UB - LB\right)\right], \ \ \ \text{if} \ \ u_5 < 0.5
\end{matrix} \right.
\label{HHO_1}
\end{equation}

\begin{equation}
x_{mean}^t = \dfrac{1}{n_{\text{part}}} \sum^{n_{\text{part}}}_{i=1} x^t_i
\label{HHO_2}
\end{equation}

To transit from exploration to exploitation the parameter $E^{\text{HHO}}$ is applied, which can be seen in Eq. \ref{HHO_3}, where $E^{\text{HHO}}$ indicates the escaping energy of the prey, $k_{max}$ is the maximum number of iterations and \(E^{\text{HHO}}_0 \in [-1, 1]\) is the initial state of its energy. From the calculation of $E^{\text{HHO}}$ the algorithm will alternate the mechanism of search. If \(|E^{\text{HHO}}| \ge 1\), exploration phase take account. If \(|E^{\text{HHO}}| < 1\), exploitation phase is considered. Besides that parameter, a random number $u_6$ is used to determine if the prey has or has not successfully escaped. Numerically, this controls besieges with progressive rapid dives, that is, more rigorous exploitation of the method.

\begin{equation}
E^{\text{HHO}} = 2 \cdotp E^{\text{HHO}}_0  \cdotp\left(1- \dfrac{t}{k_{max}}\right)
\label{HHO_3}
\end{equation}

If \(u_6 \ge 0.5\) and \(|E^{\text{HHO}}| \ge 0.5\), Eqs. \ref{HHO_4} and \ref{HHO_5} will be used for ``soft besiege", where \(\Delta x_i^t\) is the difference between the position of the rabbit and the current location in iteration $t$ and \(J^{\text{HHO}} = 2 \cdotp (1 - u_6)\) represents the random jump strength of the rabbit throughout the escaping procedure. The $J^{\text{HHO}}$ value changes randomly in each iteration to simulate the nature of rabbit motions. 

\begin{equation}
x^{t+1}_i = \Delta x^t_i - E^{\text{HHO}} \cdotp |J^{\text{HHO}} \cdotp x_*^t - x^t_i|
\label{HHO_4}
\end{equation}

\begin{equation}
\Delta x^t_i = |x^t_* - x^t_i|
\label{HHO_5}
\end{equation}

When \(u_6 \ge 0.5\) and \(|E^{\text{HHO}}| < 0.5\), Eq. \ref{HHO_6} is used. This mechanism is called ``hard besiege".

\begin{equation}
x^{t+1}_i = x_*^t - E^{\text{HHO}} \cdotp |\Delta x_i^t|
\label{HHO_6}
\end{equation}

For ``soft besiege with progressive rapid dives", that is \(u_6 < 0.5\) and \(|E^{\text{HHO}}| \ge 0.5\), Eqs. \ref{HHO_7}, \ref{HHO_8}, \ref{HHO_9}, \ref{HHO_9novo} and \ref{HHO_10} are used. In these equations \(L_{\text{HHO}}\) is the L\'evy flight function. For L\'evy flight computation, two Gaussian distributions are used, \(n_g = \text{N}\left(0, \sigma_{\text{HHO}}\right)\) and \(n_1 = \text{N}\left(0, 1\right)\), \(\lambda_{\text{HHO}}\) is a default constant and \(\Gamma\) represents the Gamma function.

\begin{equation}
y_i^t = x^t_* - E^{\text{HHO}} \cdotp |J^{\text{HHO}} \cdotp x^t_* - x^t_{i}|
\label{HHO_7}
\end{equation}

\begin{equation}
z^t_i = y^t_i +  u_7 \cdotp 
L_{\text{HHO}} \left(\lambda_{\text{HHO}}\right)\label{HHO_8}
\end{equation}

\begin{equation}
L_{\text{HHO}} \left(\lambda_{\text{HHO}}\right) = 0.01 \cdotp \dfrac{n_g \cdotp \sigma_{\text{HHO}}}{|n_1|^{\frac{1}{\lambda_{\text{HHO}}}}} \label{HHO_9}
\end{equation}

\begin{equation}
\sigma_{\text{HHO}} = \left[\dfrac{\Gamma\left(1+\lambda_{\text{HHO}}\right)}{\lambda_{\text{HHO}} \cdotp \Gamma \left(\frac{1 + \lambda_{\text{HHO}}}{2}\right)} \cdotp \dfrac{\text{sin}\left(\frac{\pi \cdotp \lambda_{\text{HHO}}}{2}\right)}{2^{\frac{\lambda_{\text{HHO}} - 1}{2}}}     \right]^{\frac{1}{\lambda_{\text{HHO}}}} \label{HHO_9novo}
\end{equation}

\begin{equation}
x^{t+1}_i =  \left\{\begin{matrix}
y_i^t, \ \ \ \text{if} \ \ \mathcal{F}\left(y_i^t\right) < \mathcal{F}\left(x_i^t\right)
\\ 
z^t_i, \ \ \ \text{if} \ \ \mathcal{F}\left(z_i^t\right) < \mathcal{F}\left(x_i^t\right)
\end{matrix} \right.
\label{HHO_10}
\end{equation}

At last, for ``hard besiege with progressive rapid dives'', that is \(u_6 < 0.5\) and \(|E^{\text{HHO}}| < 0.5\), Eq. \ref{HHO_12} and Eq. \ref{HHO_13} can be used to calculate the next position by Eq. \ref{HHO_11}.

\begin{equation}
y_i^t = x^t_* - E^{\text{HHO}} \cdotp |J^{\text{HHO}} \cdotp x^t_* - x^t_{mean}|
\label{HHO_12}
\end{equation}

\begin{equation}
z^t_i = y^t_i +  u_7 \cdotp 
L_{\text{HHO}} \left(\lambda_{\text{HHO}}\right)
\label{HHO_13}
\end{equation}

\begin{equation}
x^{t+1}_i =  \left\{\begin{matrix}
y_i^t, \ \ \ \text{if} \ \ \mathcal{F}\left(y_i^t\right) < \mathcal{F}\left(x_i^t\right)
\\ 
z^t_i, \ \ \ \text{if} \ \ \mathcal{F}\left(z_i^t\right) < \mathcal{F}\left(x_i^t\right)
\end{matrix} \right.
\label{HHO_11}
\end{equation}

A more detailed view of the iterative process considered in this method can be seen on Algorithm \ref{alg_HHO}.

\begin{algorithm} [H]
	\caption{Pseudo-code of HHO algorithm \cite{heidari2019harris}}
	\begin{algorithmic}
		\State Initialize the random population $\mathbf{X}$
		\While{(criterion)}
		\State Calculate the fitness values of hawks
		\State Set \textbf{$x^t_*$} as the location of rabbit (best location)
		\For{earch search agent}
		\State Update the initial energy $E^{\text{HHO}}_{0}$ and jump strength $J^{\text{HHO}}$
		\State Update the $E^{\text{HHO}}$ using Eq. \ref{HHO_3}
		\If{ ($|E^{\text{HHO}}|\geq 1$)} 
		\State Update location using Eq. \ref{HHO_1}
		\EndIf
		\State \textbf{end if}
		\If {($|E^{\text{HHO}}|< 1$)} 
		\If{($u_6\geq$ 0.5 and $|E^{\text{HHO}}|\geq 0.5$ ) }  
		\State Update using Eq. \ref{HHO_4}
		\ElsIf{($u_6\geq$ 0.5 and $|E^{\text{HHO}}|< 0.5$ ) } 
		\State Update using Eq. \ref{HHO_6}
		\ElsIf{($u_6<$0.5 and $|E^{\text{HHO}}|\geq 0.5$ ) } 
		\State Update using Eq. \ref{HHO_10}
		\ElsIf{($u_6<$0.5 and $|E^{\text{HHO}}|< 0.5$ ) } 
		\State Update using Eq. \ref{HHO_11}
		\EndIf
		\State \textbf{end if}
		\EndIf
		\State \textbf{end if}
		\EndFor 
		\State \textbf{end for}
		\EndWhile
		\State \textbf{end while}
		\State Return \textbf{$x_*^t$}
	\end{algorithmic}
	\label{alg_HHO}
\end{algorithm}

\section{Markov Chain Monte Carlo}
\label{mcmc}

Markov Chain Monte Carlo (MCMC) is an statistical approach that consists in three methods: Monte Carlo, Markov Chain and, in this paper, Metropolis-Hasting algorithm. Monte Carlo refers to methods that are based on the generation of random numbers from a distribution. Mathematically, the Monte Carlo method can be expressed as a random walk with a normal distribution given by \( \theta_t = \mu_{\theta} \left( 1 + \sigma^2_{\theta} \xi \right )   \), where \( \theta_t \) represents the parameter under analysis, \(\mu_{\theta}\) and \(\sigma^2_{\theta}\), respectively, the mean and variance of \(\theta\) and \(\xi\) is a random variable with zero mean. Here, we assumed $\xi_i \in [-1,1] $ with uniform distribution and \(\theta \in \Theta\).  

A Markov Chain is a stochastic process \(\left \{\theta_0, \theta_1, \theta_2, ... \right \} \) such that the distribution of \(\theta_i\)  given all previous values $ \theta $ depends only on the immediately preceding $ \theta $, that is, \(\theta_{i-1} \). Mathematically, in Eq. \ref {eq_MC}, is observed that,
\begin{flalign}
& \mathbf{P} \left ( \theta_i \in R| \theta_0,...,\theta_{i-1} \right ) = \mathbf{P} \left ( \theta_i \in R| \theta_{i-1} \right ) \label{eq_MC}
\end{flalign}
for any subset \textit{R}.

Thus, MCMC is an iterative version of the usual Monte Carlo method. From a bayesian view, the solution of an inverse problem is a probability density function of $\theta$ given the observations \(X\) a posteriori \(\mathbf{P}\left (\theta|X\right)\), where \(X\) represents the data vector and $\theta$ the parameter vector. The probability density function of $\theta$, given \(X\), can be written according to Bayes rule as Eq. \ref{eq_bayes},
\begin{flalign}
& \mathbf{P} \left ( \theta | X \right ) = \dfrac{\mathbf{P} \left ( X | \theta \right ) \cdotp \mathbf{P} \left ( \theta \right )}{\mathbf{P} \left ( X \right )}
\label{eq_bayes}
\end{flalign}
where the likelihood is \(\mathbf{P}\left(X|\theta\right)\), \(\mathbf{P} \left (\theta \right) \) is the a priori distribution of the parameters and \( \mathbf{P} \left( X \right) =  \int_{\Theta}  \mathbf{P} \left ( X | \theta \right ) \cdotp \mathbf{P} \left ( \theta \right ) d\theta \) is a normalizing factor \cite{chen2003bayesian}. However, once \(\mathbf{P} \left( X \right) \) is a constant, this information may be temporarily disregarded in order to obtain a posteriori probability density function as Eq. \ref {eq_bayes2},
\begin{flalign}
& \mathbf{P} \left ( \theta | X \right ) \propto  \mathbf{P} \left ( X | \theta \right ) \cdotp \mathbf{P} \left ( \theta \right )
\label{eq_bayes2}
\end{flalign}

Considering that the parameters are linearly independent, equally distributed and presents Gaussian probability density, \(\mathbf{P} \left ( \theta \right )\) can be modeled as shown in Eq. \ref {eq_priori}.

\begin{flalign}
& \mathbf{P} \left ( \theta \right ) \propto  \text{exp} \left [-\dfrac{1}{2} \cdotp\left (\dfrac{\theta - \mu_{\theta}}{\sigma_{\theta}} \right)^2 \right ]
\label{eq_priori}
\end{flalign}
where $\theta$ is the parameter, $\mu$ is the average of the Gaussian distribution and $\sigma_{\theta}$ is the Gaussian priori standard deviation. The priori is the product of the prioris at each point, that is, by Eq. \ref {eq_priori_produtorio},

\begin{flalign}
& \mathbf{P} \left ( \theta \right ) = \prod_{i=1}^{\text{Nvar}}  \mathbf{P} \left ( \theta_i \right )
\label{eq_priori_produtorio}
\end{flalign}
where \(\text{N}_{\text{var}}\) is the number of state variables. Thus, using the Eqs. \ref {eq_priori} and \ref {eq_priori_produtorio}:
\begin{flalign}
& \mathbf{P} \left ( \theta \right ) =  \text{exp} \left [-\dfrac{1}{2} \cdotp \sum_{i=1}^{\text{Nvar}} \left (\dfrac{\theta_i - \mu_{\theta_i}}{\sigma_{\theta_i}} \right)^2 \right ]
\label{eq_priori_condensada}
\end{flalign}

Similarly, likelihood is proportional to an exponential function, as stated in Eq. \ref {eq_Lk}, and total likelihood is equal to the likelihood output of each parameter, as indicated in Eq. \ref {eq_Lk_produtorio},
\begin{flalign}
& \mathbf{P} \left ( X |\theta \right ) \propto  \text{exp} \left \{-\dfrac{1}{2} \left [\dfrac{\left ( X^* - X(\theta_i) \right ) \cdotp \left ( X^* - X(\theta_i) \right )^T }{\sigma_i^2} \right] \right \}
\label{eq_Lk}
\end{flalign}
\begin{flalign}
& \mathbf{P} \left ( X |\theta \right ) = \prod_{i=1}^{\text{Nvar}}  \mathbf{P} \left ( X |\theta_i \right )
\label{eq_Lk_produtorio}
\end{flalign}
where \(X^*\) are the experimental data and \(X \left( \theta\right) \) is the state variable value in $\theta$. So, as seen in Eq. \ref{eq_Lk_condensada}, the likelihood can be expressed as:
\begin{flalign}
\mathbf{P} \left ( X |\theta \right ) =  \text{exp} \left \{-\dfrac{1}{2}  \sum_{i=1}^{\text{Nvar}} \left [ \left (\dfrac{\left ( X^* - X(\theta_i) \right ) \left ( X^* - X(\theta_i) \right )^T }{\sigma^2_{\theta_i}} \right) \right ]\right \}
\label{eq_Lk_condensada}
\end{flalign}

The Metropolis-Hasting algorithm is used to decide which values to accept or discard. We begin by calculating the later probability of the new parameter and the previously accepted parameter, as stated in Eq. \ref {eq_MH_1}.

\begin{flalign}
& \alpha\left(\theta,\theta^* \right ) = \text{min} \left[ 1,\dfrac{\mathbf{P}\left(\theta|X\right)^*}{\mathbf{P}\left(\theta|X\right)}\right] = \text{min} \left[ 1,\dfrac{\mathbf{P}\left(X|\theta^*\right) \mathbf{P}\left(\theta^* \right )}{\mathbf{P}\left(X|\theta\right) \mathbf{P}\left(\theta \right )}\right]
\label{eq_MH_1}
\end{flalign}

This method will be used to optimize the interface parameters of the temporal model exposed in Eq. \ref{atb}. After obtaining the uncertainty interval of these parameters, it will be considered as reference data to evaluate the performance of the metaheuristic methods.

\section{Inverse Methodology}
\label{methodology}

The experimental data related of the thermal profile of Al-7wt.\%Si alloy, as well as thermophysical properties considered in this paper, can be found in \cite{PeresTese}. For performance analysis of the metaheuristic algorithms  presented in Section \ref{metaheuristics}, the following settings have been standardized: 

\begin{itemize}
	\item Search range of parameter \(A\) : [0, 10000] \(\frac{\text{W}}{\text{m}^2 \text{K}}\);
	\item Search range of parameter \(B\): [-0.5, -0.005];
	\item Maximum number of iterations on each optimization: 100;
	\item Number of particles: 20;
	\item Stop criterion: 10 iterations resulting on the same best parameter particle.
\end{itemize}

Table \ref{tab_param_meta} contains the parameters intrinsic to each algorithm considered in this research.

\begin{table}[H]
	\caption{Parameter values used on the metaheuristic algorithms.}
	\centering
	\resizebox{12cm}{!}{%
		\begin{tabular}{|c|c|c|}
			\hline
			Method & Parameters & Reference \\ \hline
			
			Particle Swarm Optimization & \begin{tabular}[c]{@{}c@{}} Acceleration constants \(\alpha_{\text{PSO}}\) and \(\beta_{\text{PSO}} = 2\) \\ Inertia constant \(\theta_{\text{PSO}}\) = 1\end{tabular} & \cite{kennedy1995particle} \\ \hline
			
			Differential Evolution & \begin{tabular}[c]{@{}c@{}} Differential weights \(F^{1}_{\text{DE}} = F^{2}_{\text{DE}} = 0.5\) \\ Crossover parameter \(C_{r_{\text{DE}}} = 0.8\)\end{tabular} & \cite{storn1997} \\ \hline
			
			Bat Algorithm & \begin{tabular}[c]{@{}c@{}} Loudness \(\varphi_{\text{BA}} = 0.25\)\\ Pulse rate \(r_{\text{BA}} = 0.5\)\\ Frequency \(f_{\text{BA}} \in \left[ 0,2  \right]\)\\ Constants $\gamma_{\text{BA}}$ and $\alpha_{\text{BA}}$ = 0.9 \\ Scaling factor \(\sigma_{\text{BA}}\) = 1 \end{tabular} & \cite{yang2010} \\ \hline
			
			Flower Pollination Algorithm & \begin{tabular}[c]{@{}c@{}}Switch probability \(p_{\text{FPA}} = 0.8\)\\ L\'evy flight parameter \(\lambda_{\text{FPA}} = 1.5\)\\ Scaling factor \(\gamma_{\text{FPA}} = 0.1\)\end{tabular} & \cite{yang2012}\\ \hline
			
			Grey Wolf Optimizer & - & \cite{mirjalili2014grey} \\ \hline
			
			Moth-Flame Optimization & Spiral shape parameter \(b_{\text{MFO}} = 1\) & \cite{mirjalili2015moth}\\ \hline
			
			Sine Cosine Algorithm & Weighting factor \(a_{\text{SCA}} = 2\) & \cite{mirjalili2016sca} \\ \hline
			
			Whale Optimization Algorithm & Spiral shape parameter \(b_{\text{WOA}} = 1\) & \cite{mirjalili2016whale} \\ \hline
			
			Dragonfly Algorithm & L\'evy flight parameter \(\lambda_{\text{DA}} = 1.5\) & \cite{mirjalili2016dragonfly}\\ \hline
			
			Harris Hawks Optimization & L\'evy flight parameter \(\lambda_{\text{HHO}} = 1.5\) & \cite{heidari2019harris}\\ \hline
		\end{tabular}%
		\label{tab_param_meta}
	}
\end{table}

\begin{figure}[htbp]
	\caption{Markov Chains initial states: Burn-in regions and evolution for the two parameters.}
	\centering
	\includegraphics[width=0.8 \linewidth]{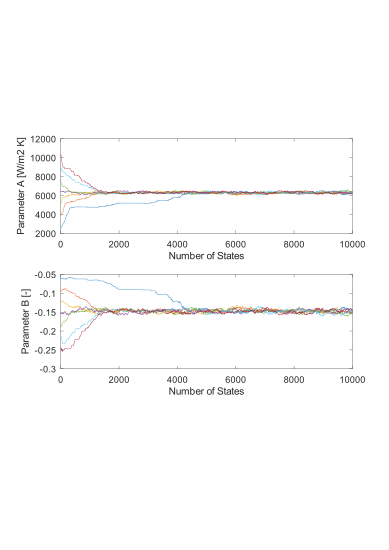}
	\label{burn-in}
\end{figure}

The metrics to be analyzed in this contribution, for metaheuristic methods, will be: Sample mean, standard deviation and kurtosis of the optimal points, average number of iterations, convergence and error analysis. Regarding the MCMC method (presented in Section \ref{mcmc}), the parameters considered were as follows:

\begin{itemize}
	\item Priori information of parameter \(A\): 6430 \(\frac{\text{W}}{\text{m}^2 \text{K}}\) \cite{edilma2019};
	\item Priori information of parameter \(B\): -0.153 \cite{edilma2019};
	\item Number of Markov Chains analyzed: 7;
	\item Markov Chain starting points: 40\%, 60\%, 80\%, 100\%, 120\%, 140\% and 160\% from the a priori information (Fig. \ref{burn-in});
	\item Number of states for each Markov Chain: 40000;
	\item Search Step: \(0.005 \times \text{N} \left(0,1\right) \times \theta_{\left(i-1\right)}\);
	\item Standard deviation of the experimental measures considered: 5 K.
\end{itemize}

\begin{figure}[htbp]
	\caption{Markov chains obtained by MCMC.}
	\centering
	\includegraphics[width=0.8 \linewidth, trim=0cm 0cm 0cm 0cm, clip=true]{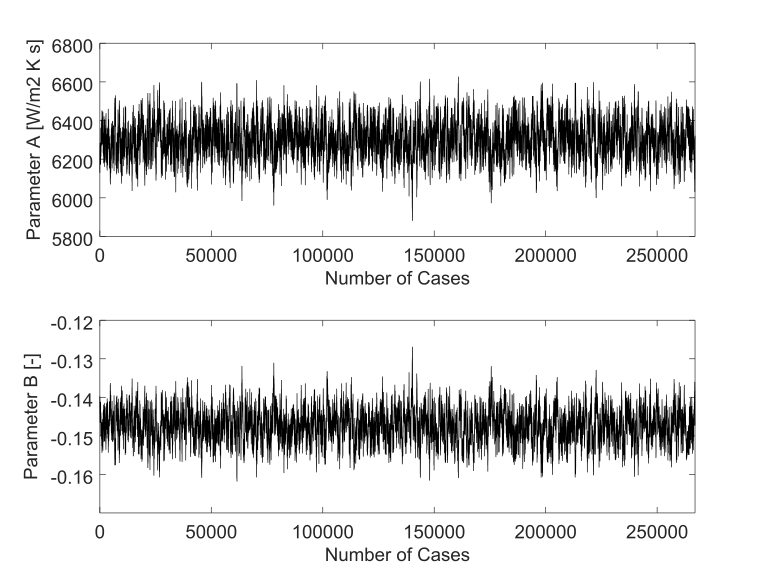}
	\label{cadeia}
\end{figure}

After removing the burn-in samples, as can be seen in Fig. \ref{burn-in}, the Markov chains were concatenated and analyzed to obtain the expected value, standard deviation, percentiles 95\% and 5\% of each parameter under analysis. The concatenation of Markov chains, resulting in 267000 states, can be seen in Fig. \ref{cadeia}.

\section{Results and Discussion}
\label{results_discussion}

\subsection{Results via Markov Chain Monte Carlo}
\label{results_MCMC}

\begin{figure}[htbp]
	\caption{Histograms obtained by MCMC.}
	\centering
	\includegraphics[width=0.75 \linewidth, trim=1cm 6.5cm 1cm 7cm, clip=true]{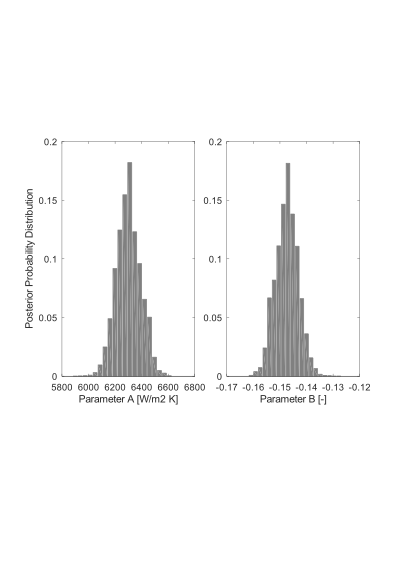}
	\label{barras_mcmc}
\end{figure}

The posterior probability distribution obtained by MCMC of parameters $A$ and $B$ can be seen in Fig. \ref{barras_mcmc}. The 267000 states are represented, in this figure, by 20 bins, based on Doane's formula \cite{doane1976aesthetic}. Presenting a positive excess-kurtosis and skewness of $3\times$10$^{-2}$ each, these histograms are leptokurtic and approximately symmetrical.

Based on Fig. \ref{barras_mcmc}, the parameter value that minimizes the model error with experimental data, ie, has the highest posterior probability value, is between 6200 and 6400 \(\frac{\text{W}}{\text{m}^2 \text{K}}\) for parameter \(A\) and between -0.15 and -0.14 for parameter \(B\). The expected value, such as percentile of 95\%, 5\% and standard deviation of the Markov Chain states are show in Tab. \ref{tab_resul_mcmc}. Here, percentiles of 95\% and 5\% are denominated, respectively, the maximum and minimum value of each parameter.

\begin{table}[htbp]
	\caption{Expected value, standard deviation, maximum and minimum values of the parameters obtained by MCMC.}
	\centering
	\resizebox{12cm}{!}{%
		\begin{tabular}{|c|c|c|c|c|}
			\hline
			Parameter & Expected Value & Standard Deviation & Minimum Value &  Maximum Value \\ \hline
			A [W/m$^2$ K] & 6301 & 91 & 6126& 6476 \\ \hline
			B [-] & -0.147 & 0.004 & -0.156  & -0.139 \\ \hline
		\end{tabular}%
	}
	\label{tab_resul_mcmc}
\end{table}

\begin{figure}[htbp]
	\caption{Experimental and simulated thermal profiles of Al-7wt.\%Si solidification.}
	\centering
	\includegraphics[width=0.7 \linewidth, trim=1cm 6.5cm 1cm 7cm, clip=true]{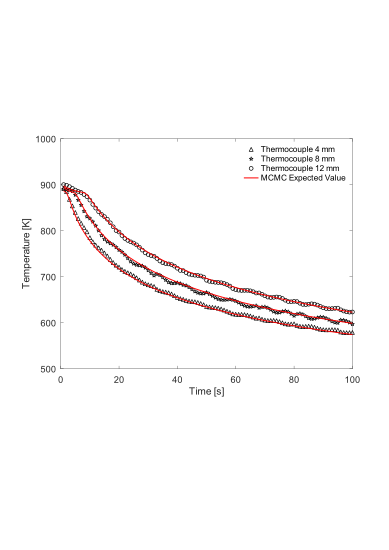}
	\label{profile}
\end{figure}

In Figure \ref{profile} is shown the numerical thermal profile resulted by the expected value of \textit{A} and \textit{B} obtained by MCMC. The small distance between the experimental and numerical curves corroborates the accuracy of the values raised for the heat exchange interface parameters. Applying the expected value obtained by MCMC in the simulatin, the standard deviation between the experimental and simulated thermal profile for the thermocouples in 4 mm, 8 mm and 12 mm were, respectively, 4.05 K, 3.72 K and 2.82 K.

\subsection{Performance Analysis of Metaheuristic Algorithms}

In this section a performance analysis of all metaheuristic algorithms expressed on Section \ref{metaheuristics} is exposed. In order to present the results more coherently, this section will have four subsections: Sample Mean, Percentiles and Kurtosis (\ref{sub:mean}), Error Analysis (\ref{sub:error}), Average Number of Iterations and Convergence (\ref{sub:conv}) and Overall Performance (\ref{sub:geral}).

An evaluation table of all the information present in this section can be seen in Tab. \ref{tab_resume}. The quantitative data were represented qualitatively based on parameters derived both from the information obtained by MCMC method and the performance of each metaheuristic algorithm. In this table, ``Relative Performance" is based on the comparison of the results between the metaheuristic methods and MCMC. ``Number of Iterations" stands for the average number of iterations of each optimizer. ``Convergence" represents how the method was able to search the best minimum on the domain without being trapped on a local one. ``Average Error" is a qualitative evaluation of the average error of the optimized values of each method under analysis. ``Overall Performance" indicates a subjective evaluation of the authors about the results available during this research, taking into account all the metrics analyzed.

\begin{table}[htbp]
	\caption{Comparative performance analysis of the metaheuristic optimizers.}
\centering
	\begin{tabular}{|c|c|c|c|c|c|}
		\hline
		Methods & \begin{tabular}[c]{@{}c@{}}Relative\\ Performance\end{tabular} & \begin{tabular}[c]{@{}c@{}}Number of\\ Iterations\end{tabular} & Convergence & \begin{tabular}[c]{@{}c@{}}Average\\ Error\end{tabular} & \begin{tabular}[c]{@{}c@{}}Overral\\ Performance\end{tabular} \\ \hline
		PSO     & Excellent      & High         & Satisfactory & Low          & Excellent                                                               \\ \hline
		DE       &  Excellent      & Medium  & Good        & Medium  & Good                                                             \\ \hline
		BA       &  Regular        & Low          & Premature   & High         & Regular                                                              \\ \hline
		FPA     &  Regular        & Low          & Premature   & High         & Regular                                                              \\ \hline
		GWO  &   Good           &  Medium  & Satisfactory  & Low          & Excellent                                                              \\ \hline
		MFO    &   Excellent    &  High        & Satisfactory & Low          & Excellent                                                              \\ \hline
		SCA     &   Regular      & Low  & Premature    & Medium  & Regular                                                              \\ \hline
		WOA   &   Regular      & Medium  & Good        & Low          & Good                                                              \\ \hline
		DA       &   Regular      & Low          & Premature     & Medium  & Regular                                                              \\ \hline
		HHO    &  Good           & Low          & Premature     & High        & Regular                                                              \\ \hline
	\end{tabular}
	\label{tab_resume}
\end{table}

\subsubsection{Sample Mean, Percentiles and Kurtosis}
\label{sub:mean}

\begin{figure}[htbp]
	\caption{Uncertainty range and kurtosis of the metaheuristic methods and MCMC, for parameter \(A\).}
	\centering
	\includegraphics[width=0.9 \linewidth, trim=1cm 6.5cm 1cm 6.2cm, clip=true]{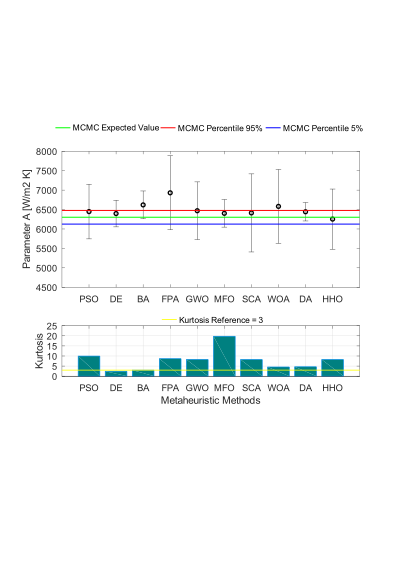}
	\label{med_desvio_a}
\end{figure}

In Figure \ref{med_desvio_a}, it is shown the quantitative results related to the studied optimizers. In this, the methods on the horizontal axis of the charts are in chronological sequence with respect to their publication. The horizontal lines represent the results obtained by MCMC. The circles represent the sample mean of the optimal values of each method, along with the standard deviation bars above and below that point. The second graphic shows the kurtosis of each method, in which the yellow horizontal line on the blue bars indicates the value 3, that guides the argument about tail size of the distribution presented in each method.

It is clear, based on the uncertainty range of each method, that all methods were able, at different coherence levels, to estimate the value of parameter \(A\), considering the uncertainty range obtained by MCMC method. It is observed that, in a range from 0 to 10000 \(\frac{\text{W}}{\text{m}^2 \text{K}}\), the results of all methods are quite consistent, indicating that any method out of ten can retrieve the value of this parameter. However, some considerations can be taken from the chart:

\begin{itemize}
	\item BA, FPA, and WOA presented a sample mean outside the uncertainty range of MCMC method. Considering the kurtosis of each method, it is observed that they do not recover exactly and precisely the expected value of MCMC. Even though the uncertainty range of each method covers part or all of MCMC uncertainty range, the probability that these methods will optimize the parameter to the exact range is quite small;
	
	\item PSO, GWO, SCA and DA presented a sample mean within MCMC uncertainty range, but exposed a wide range compared to the other methods, covering a region of approximately 2000 \(\frac{\text{W}}{\text{m}^2 \text{K}}\), or a small range that not cover all MCMC uncertainty range;
	
	\item Among the analyzed methods, DE, MFO and HHO stand out. The sample mean of these methods is quite consistent with the reference range and the standard deviation comprises a small range of values (even thought HHO does not presents a small range, its sample mean is very satisfactory, combined with a good kurtosis value).	It is noteworthy the ability of MFO to recover, in a small uncertainty range and very high kurtosis, the optimal value of parameter \(A\).
\end{itemize}

\begin{figure}[H]
	\caption{Uncertainty range and kurtosis of the metaheuristic methods and MCMC, for parameter \(B\).}
	\centering
	\includegraphics[width=0.9 \linewidth, trim=1cm 6.5cm 1cm 6cm, clip=true]{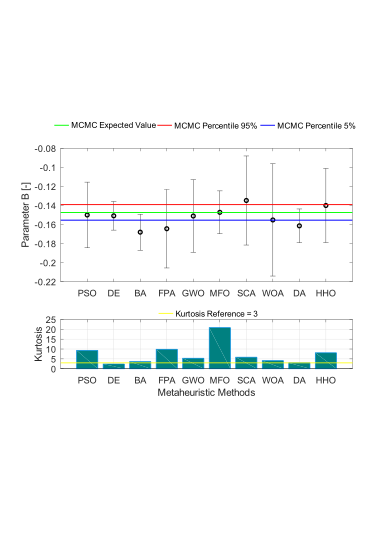}
	\label{med_desvio_b}
\end{figure}

In Figure \ref{med_desvio_b} is expressed the same quantitative parameters of statistical analysis as shown in Fig. \ref{med_desvio_a}, but relative to parameter \(B\).

As expressed on parameter $A$ analysis, at different levels, all methods were able to estimate the value of parameter \(B\), considering the MCMC uncertainty range. In the interval from -0.005 to -0.5, the results of all methods are consistent. However, it can be seen that the methods had greater difficulty in optimizing parameter $B$, which can be seen in the figure above. From the chart it is possible to visualize that:

\begin{itemize}
	\item BA, FPA, SCA, WOA and DA presented a sample mean out of the uncertainty range of MCMC method. Even if WOA has a sample mean within the uncertainty range, the standard deviation presented is so large that it does not characterize the method satisfactorily.

	\item HHO presented a sample mean almost within MCMC uncertainty range, but exposed a very wide range compared to other methods (but not so wide as WOA);
	
	\item PSO, DE, GWO and MFO stand out. The consistency of the sample mean in relation to the reference range and the small standard deviation corroborate the excellent result of these methods. DE did not present high kurtosis value, but its performance is satisfactory. It is noteworthy the performance of MFO, which presented an almost exact sample mean and reaches a kurtosis value above 20. This indicates that MFO, besides being accurate, is extremely precise.
\end{itemize}

\subsubsection{Error Analysis}
\label{sub:error}

In this subsection, the optimization errors of the metaheuristic algorithms under study will be analyzed. Consider, by notation, that 
$\mathcal{F}\left(\Box\right)$ is the fitness function and $\mathbf{G}_*$ is the vector that corresponds to the optimized values of each method. $\mathcal{F}\left(\mathbf{G}_*\right)$ stands for the error of each optimized parameter vector. Considering that we are evaluating three thermocouple positions and the error parameter is the standard deviation between experimental and numerical thermal profiles \(\left(\sigma^{\text{therm}}_{i}\right) \), $\mathcal{F}\left(\mathbf{G}_*\right)$ can be represented as $\mathcal{F}\left(\mathbf{G}_*\right) = \sum_{1}^{3}\sigma^{\text{therm}}_i$.

Five metrics will be considered in terms of error: absolute sum  $\left(\sum\mathcal{F}\left(\mathbf{G}_*\right) \right)$, expected value $\left(\text{E} \left[ \mathcal{F}\left(\mathbf{G}_*\right) \right] \right)$, standard deviation $\left(\sigma\left(\mathcal{F}\left(\mathbf{G}_*\right) \right) \right)$ and maximum  $\left(\text{max}\left(\mathcal{F}\left(\mathbf{G}_*\right) \right)\right)$ and minimum  $\left(\text{min}\left(\mathcal{F}\left(\mathbf{G}_*\right) \right)\right)$ values. In Table \ref{tab_error_meta} is shown the abovementioned metrics for each metaheuristic method analyzed.

\begin{table}[htbp]
		\caption{Absolute sum, expected value, standard deviation, maximum and minimum values of error for the metaheuristic algorithms under analysis.}
	\centering
	\begin{tabular}{|c|c|c|c|c|c|}
		\hline
		Methods & $\sum\mathcal{F}\left(\mathbf{G}_*\right) $ &$\text{E} \left[ \mathcal{F}\left(\mathbf{G}_*\right) \right] $ & $\sigma\left(\mathcal{F}\left(\mathbf{G}_*\right) \right) $ & $\text{max}\left(\mathcal{F}\left(\mathbf{G}_*\right) \right)$ & $\text{min}\left(\mathcal{F}\left(\mathbf{G}_*\right) \right)$ \\ \hline
		PSO   &418.78  & 10.47    & 0.34         & 12.07   & \textbf{10.31}   \\ \hline
		DE     & 447.49  & 11.19    & 0.79        & 13.40   & 10.34   \\ \hline
		BA     &1030.30 & 25.76    & 18.05      & 78.89   & 10.66   \\ \hline
		FPA   &960.91  & 24.02    & 16.15       & 84.36   & 10.80   \\ \hline
		GWO &420.71  & 10.52    & 0.15        & 10.91   & 10.32   \\ \hline
		MFO  &\textbf{413.71}  & \textbf{ 10.34}    & \textbf{0.05}       & \textbf{10.60}   & \textbf{10.31}   \\ \hline
		SCA   &469.8    & 11.75    & 1.31        & 14.75   & 10.41   \\ \hline
		WOA &428.56  &  10.71    & 0.61       & 13.19   & 10.32   \\ \hline
		DA      & 457.57& 11.44    & 1.55        & 17.14   & 10.39   \\ \hline
		HHO  &490.56 & 12.26    & 3.26        & 29.41   & 10.42   \\ \hline
	\end{tabular}

	\label{tab_error_meta}
\end{table}

Considering only $\sum\mathcal{F}\left(\mathbf{G}_*\right) $ and $\text{E} \left[ \mathcal{F}\left(\mathbf{G}_*\right) \right] $, it is possible to comment that:

\begin{itemize}
	\item MFO, PSO, GWO and WOA, in this order, presented very good results, outperforming the other methods in these metrics. The data show that, on average, the optimization of these methods was quite satisfactory, reaching low error values and presenting a very low $\text{E} \left[ \mathcal{F}\left(\mathbf{G}_*\right) \right] $, sometimes being less than the error obtained by the expected value of MCMC method;
	\item DE showed good results, such as low accumulated error and expected value. However, its expected value was almost 10\% higher than the best result of this parameter (MFO), reducing its relative efficiency;
	\item DA, SCA, HHO, FPA and BA, in that order, presented unsatisfactory results in these metrics. In comparison with better performance methods, these presented very high values of accumulated error and expected error values of 1.1 to 2.5 times greater than the lowest expected value and accumulated error found (MFO).
\end{itemize}

Based on $\sigma\left(\mathcal{F}\left(\mathbf{G}_*\right) \right) $, $\text{max}\left(\mathcal{F}\left(\mathbf{G}_*\right) \right)$ and $\text{min}\left(\mathcal{F}\left(\mathbf{G}_*\right) \right)$, it is possible to argue that:

\begin{itemize}
	\item  MFO and GWO presented very small maximum intervals, indicating high precision. The two methods remained in the error range 10 to 11, providing the best results compared to all methods. The small standard deviation value presented by MFO is highlighted, which, corroborated by the high kurtosis presented in subsection \ref{sub:mean}, indicates that this method is the most accurate and precise of all;
	\item PSO and WOA showed good results such as small maximum interval and standard deviation. However, the maximum value presented by the two methods is at least 10 \% higher than the lowest maximum values found (MFO);
	\item DE, SCA, DA, HHO, BA and FPA did not present satisfactory results. In addition to high standard deviation, their maximum value are 1.25 to 8 times greater than the best maximum value (MFO).
\end{itemize}

It is important to note that, due to the results of $\text{min}\left(\mathcal{F}\left(\mathbf{G}_*\right) \right)$, it is observed that at least in one optimization, all methods were able to satisfactorily minimize the parameters of the IHTC under study.

\subsubsection{Average Number of Iterations and Convergence}
\label{sub:conv}

\begin{figure}[H]
	\caption{Average number of iterations for each optimization method.}
	\centering
	\includegraphics[width=0.65 \linewidth, trim=1cm 6.5cm 1cm 7cm, clip=true]{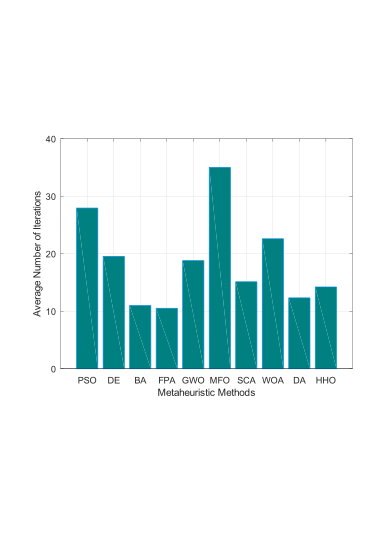}
	\label{iter}
\end{figure}

In Figure \ref{iter} is shown the average number of iterations of each method until convergence. As stated in the Section \ref{methodology}, the stopping criterion considered in this paper is the repetition of the same parameter vector for 10 iterations.

 It is possible to note from results that:
\begin{itemize}
	\item BA, FPA, SCA, DA and HHO presented low average number of iterations, since the average of iterations were very close to the minimum stopping criterion. Given their poor performance shown in Figs. \ref{med_desvio_a} and \ref{med_desvio_b}, it is possible to assume that this regular performance in parameter optimization indicates premature convergence;
	\item DE, GWO and WOA presented an average number of iterations twice greater than the stopping criterion.This is not necessarily a good thing. The price for the largest number of iterations should be rewarded with increased efficiency in finding a better critical point. In this case, the DE and WOA methods were good, with useful results. However, GWO showed a better iteration/result ratio among the three, with satisfactory convergence;
	\item Finally, PSO and MFO presented an average of iterations three times higher than the stopping criterion. This indicates that the search space was better explored. This actually corroborates the quantitative results. Even though the method required a higher average number of iterations, its results were quite satisfactory.
\end{itemize}

\subsubsection{Overall Performance}
\label{sub:geral}

From the previous analysis it is possible to draw conclusions about the parameter estimation in each metaheuristic method studied. One should keep in mind that the reference values used in this study are relative to MCMC results presented on Subsection \ref{results_MCMC} and .

Particle Swarm Optimization (PSO) presented excellent relative performance. The sample mean of parameters $A$ and $B$ was into the uncertainty range of reference and presented one of the highest excess-kurtosis for both parameters. PSO's average number of iterations was high but it converged satisfactorily. In terms of error analysis, it presented the lowest minimum value of all methods and very good results, such as low expected value, accumulated error and standard deviation. It can be said, based on the above, that this method was one of the best in optimizing the IHTC parameters, presenting excellent overall performance.

Differential Evolution (DE) exposed excellent relative performance in both parameters under analysis. The uncertainty range produced by the method was consistent with the reference, indicating accuracy and precision. The average number of interactions was reasonable, indicating that the method did not show premature convergence. However, in the error analysis, it did not obtain as satisfactory results as other methods. In general, the method showed acceptable convergence and results. Thus, DE represents a good method to optimize this kind of physical problem.

Bat Algorithm (BA) showed regular relative performance in both parameters under analysis. The uncertainty interval produced by the method was quite inconsistent with the reference. This may be due to a very premature convergence, based on the average number of iterations, which has a value very close to the minimum number of acceptable iterations. This is reflected in the error analysis, which presented very bad metrics, both in accumulated error and expected value. Therefore, BA has not behaved as an adequate method for this type of inverse estimate in this case, showing regular overall performance.

Flower Pollination Algorithm (FPA) also presented regular relative performance in both parameters under analysis. The uncertainty range presented was quite inconsistent with the reference. This method, as BA, suffered very premature convergence, presenting a low number of iterations. In view of this, the error data presented by this method were not satisfactory. This presented a very high accumulated error, as well as expected value. That said, FPA had a regular overall performance.

Grey Wolf Optimizer (GWO) showed good relative performance. The uncertainty interval presented was consistent with the reference. The average number of iterations of this method was median, however it presented satisfactory convergence. This is demonstrated by the error analysis, which presented a low expected value and accumulated error. This presented one of the smallest maximum intervals among all methods, also standing out for the small standard deviation. Given this, GWO was one of the best methods analyzed, presenting excellent overall performance.

Moth-Flame Optimization (MFO) exposed excellent relative performance. The uncertainty range presented was quite consistent with the reference, both due to the proximity of the sample mean to the reference average and the small standard deviation presented. This fact is accentuated by the kurtosis of the distribution of the optimized parameters, higher than all the other methods. The average number of iterations of this method was high, however it presented very satisfactory convergence. In error analysis, MFO stood out. All the metrics presented by this model were the best, managing to be able to estimate as well, or even better, than the reference method itself. This presented the smallest maximum interval among all methods, also standing out for the small standard deviation. Given this, MFO was the best method analyzed, presenting a very satisfactory overall performance.

Sine Cosine Algorithm (SCA) presented regular relative performance. Its parameter uncertainty interval did not stand out when compared with the reference. The average number of iterations of this method was low and it showed premature convergence. When observing the performance through the error analysis, the algorithm presents median results, not as satisfactory as other methods analyzed. Based on the above, SCA presented itself as a regular method for this type of application.

Whale Optimization Algorithm (WOA) showed regular relative performance. The uncertainty range presented was quite bad compared to the reference. The standard deviation of the parameter estimate was one of the highest of all methods. The average number of iterations of this method was relatively high but presented acceptable convergence. By the error analysis, this method was satisfactory, presenting low average error in several metrics. Thus, WOA presented itself as a good method in this specific case.

Dragonfly Algorithm (DA) exposed regular relative performance. The uncertainty interval presented by the method, in both parameters under analysis, did not contain the reference interval. The average number of iterations of this method was small, which resulted in premature convergence and somewhat high error metrics. In view of these characteristics, DA did not present good results in general, being classified at the end as a regular method.

Harris Hawks Optimization (HHO) presented good relative performance. The uncertainty range exposed by the method was acceptable, with sample mean within the reference range and standard deviation, even though sharp, but with relative coherence with the reference range. However, the average number of iterations of this method was small, which resulted in premature convergence and high error metrics. In view of these characteristics, HHO did not present good results in comparison with the other methods, presenting itself as a regular method in this optimization application.

It is worth pointing out that some metaheuristic methods studied had difficulty in optimizing the parameters contained in the IHTC because the experimental data considered in this study are not smooth, presenting disturbances that create several local minimum near the optimal point. This enables premature convergence by creating dense local minimum zones. Therefore, the behavior of the metaheuristic methods considered in this paper may change when optimizing the same physical problem, but with smoother experimental data.

\section{Conclusions}
\label{conclusion}

This paper presented a qualitative and quantitative performance analysis of ten nature-inspired metaheuristic algorithms in order to verify which metaheuristic method excels in the optimization of the Interfacial Heat Transfer Coefficient (IHTC) parameters in a unidirectional permanent mold casting process for Al-7wt.\%Si alloy. For this purpose, was selected the optimizers: Particle Swarm Optimization (PSO), Differencial Evolution (DE), Bat Algorithm (BA), Flower Pollination Algorithm (FPA), Grey Wolf Optimizer (GWO), Moth-Flame Optimization (MFO), Sine Cosine Algorithm (SCA), Whale Optimization Algorithm (WOA), Dragonfly Algorithm (DA) and Harris Hawks Optimization (HHO). For that, a numerical discretization based on Finite Volume Method of the energy conservation equation in its integral form was taken into account \cite{edilma2019}. It was considered, as uncertainty range of reference, simulated data extracted from the Markov Chain Monte Carlo (MCMC) method, from 267000 number of states, taken from 7 Markov chains initiated from different starting vectors. The considered IHTC in this analysis is the temporal function \( A \left( t/t_0 \right)^{B} \), where \(A\) [\(\frac{\text{W}}{\text{m}^2 \text{K}}\)] and \(B\) [-] are constants, \(t\) represents time [s] and \(t_0\) a referential time [s] (here \(t_0 = 1 s\)).
The priori information of these methods were collected on \cite{edilma2019}. This performance analysis considered as parameters for each metaheuristic method: Expected value, standard deviation and kurtosis (for parameters $A$ and $B$), average number of iterations, convergence and absolute sum, expected value, standard deviation, maximum and minimum value of the errors between numerical and experimental thermal profiles.

From the posterior probability distribution of MCMC, based on percentile 5 \%, expected value and percentile 95 \%, the optimal uncertainty range for parameter \(A\) is, respectively, 6126 \(\frac{\text{W}}{\text{m}^2 \text{K}}\), 6301 \(\frac{\text{W}}{\text{m}^2 \text{K}}\) and 6476 \(\frac{\text{W}}{\text{m}^2 \text{K}}\) and, for parameter $B$, -0.156, -0.147 and -0.139. This range of uncertainty covers the a priori information collected by \cite{edilma2019}, validating it statistically.

Among the results obtained from the metaheuristic methods, it can be said that all of them would be able to reasonably optimize parameters \(A\) and \(B\). However, based on qualitative and quantitative metrics, BA, FPA SCA, DA and HHO did not present significant results. In general, these methods showed premature convergence, low correlation of the optimized values with the reference uncertainty range and high error values related to the optimized parameters. Thus, it can be said that these methods are not the most suitable for optimizing parameters in this specific problem.

On the other hand, DE and WOA showed reasonably good results, with adequate convergence, good average number of iterations and acceptable error metrics. They presented significantly superior results in relation to the aforementioned methods, however they did not provide a satisfactory level of error. Thus, it can be said that these methods provide good results when applied to this type of physical problem, however they are not the most suitable.

Unlike the others, PSO, GWO and MFO exposed promising results. These three presented the best quantitative results, well above the average of the previous methods. In general, the sample mean of these methods was very close to the reference expected value, their standard deviation was quite small when compared to the others, and kurtosis was quite high. The convergence of these three methods was satisfactory, with a reasonable average number of iterations. Their level of error were quite satisfactory, classifying them as excellent methods for this type of problem and application. It is important to note that MFO was the one that best optimized the parameters and reduced the uncertainty. This algorithm presented a very reduced uncertainty range, as well as the greater kurtosis. MFO was superior to the other methods in all error metrics, standing out even in comparison with the reference range itself. This fact shows that MFO is very accurated, presenting about 80\% of the optimized points within the uncertainty range resulting from MCMC.

As PSO, GWO and especially MFO outperformed the other methods in different quantitative parameters, they have total potential to optimize the IHTC parameters \(A\) and \(B\) coupled in the solidification phenomenon applied in the permanent metal mold casting. Then, all of them can be used without loss of accuracy and precision. Finally, it is worthy pointing out that the performance of all the methods analyzed in this article is extremely influenced by the values of the constants intrinsic to the numerical methodology. Thus, research on the best values to be used in the constants is encouraged in order to expand the applicability and efficiency of numerical methods in engineering and metallurgy applications such as the one studied in this contribution.

%
%
%

\bibliographystyle{abbrv}       
\bibliography{bibfile}   

\end{document}